\documentclass[sn-mathphys,Numbered]{sn-jnl}

\usepackage{graphicx}%
\usepackage{multirow}%
\usepackage{amsmath,amssymb,amsfonts}%
\usepackage{amsthm}%
\usepackage{mathrsfs}%
\usepackage[title]{appendix}%
\usepackage{xcolor}%
\usepackage{textcomp}%
\usepackage{manyfoot}%
\usepackage{booktabs}%
\usepackage{algorithm}%
\usepackage{algorithmicx}%
\usepackage{algpseudocode}%
\usepackage{listings}%
\usepackage{svg}%
\usepackage{glossaries-extra}
\usepackage{siunitx}
\usepackage[version=4]{mhchem}
\usepackage{subcaption}
\usepackage{tikz}
\usepackage[capitalise]{cleveref}
\usepackage{minitoc}
\noptcrule

\glsdisablehyper

\newacronym{Alex}{Alexandria}{Alexandria}
\newacronym{Alex-MP-ICSD}{Alex-MP-ICSD}{Alexandria - Materials Project - Inorganic Crystal Structure Database}
\newacronym{Alex-MP-20}{Alex-MP-20}{Alexandria - Materials Project structures with 20 or fewer atoms}
\newabbreviation[\glslongpluralkey={discrete denoising diffusion probabilistic models}]{D3PM}{D3PM}{discrete denoising diffusion probabilistic model}
\newabbreviation[\glslongpluralkey={denoising diffusion probabilistic models}]{DDPM}{DDPM}{denoising diffusion probabilistic model}
\newabbreviation{DFT}{DFT}{density functional theory}
\newabbreviation{DSM}{DSM}{denoising score matching}
\newabbreviation{GNN}{GNN}{graph neural network}
\newabbreviation{HHI}{HHI}{Herfindahl–Hirschman index}
\newabbreviation{ICSD}{ICSD}{Inorganic Crystal Structure Database}
\newabbreviation{MAE}{MAE}{mean absolute error}
\newabbreviation{MLP}{MLP}{multi-layer perceptron}
\newabbreviation{MP}{MP}{Materials Project}
\newacronym{MP-20}{MP-20}{Materials Project structures with 20 or fewer atoms}
\newabbreviation[\glslongpluralkey={machine learning force fields}]{MLFF}{MLFF}{machine learning force field}
\newabbreviation{PBE}{PBE}{Perdew–Burke–Ernzerhof}
\newabbreviation{RMSD}{RMSD}{root mean squared displacement}
\newabbreviation{RSS}{RSS}{random structure search}
\newabbreviation{SUN}{S.U.N.}{stable, unique, and novel}
\newabbreviation{VASP}{VASP}{Vienna ab initio simulation package}

 \DeclareSIUnit\angstrom{\text {Å}} 

\usepackage{amsmath,amsfonts,bm}

\def\eqref#1{equation~\ref{#1}}

\def\1{\bm{1}}

\def\vzero{{\bm{0}}}

\def\vmu{{\bm{\mu}}}
\def\vtheta{{\bm{\theta}}}
\def\vsigma{{\bm{\sigma}}}
\def\va{{\bm{a}}}

\def\vd{{\bm{d}}}

\def\vg{{\bm{g}}}

\def\vk{{\bm{k}}}
\def\vl{{\bm{l}}}
\def\vm{{\bm{m}}}

\def\vp{{\bm{p}}}

\def\vs{{\bm{s}}}
\def\vt{{\bm{t}}}

\def\vv{{\bm{v}}}

\def\vx{{\bm{x}}}

\def\vz{{\bm{z}}}

\def\mA{{\bm{A}}}
\def\mB{{\bm{B}}}
\def\mC{{\bm{C}}}
\def\mD{{\bm{D}}}

\def\mH{{\bm{H}}}
\def\mI{{\bm{I}}}

\def\mL{{\bm{L}}}
\def\mM{{\bm{M}}}

\def\mP{{\bm{P}}}
\def\mQ{{\bm{Q}}}
\def\mR{{\bm{R}}}

\def\mU{{\bm{U}}}
\def\mV{{\bm{V}}}
\def\mW{{\bm{W}}}
\def\mX{{\bm{X}}}

\def\mPhi{{\bm{\Phi}}}

\def\mSigma{{\bm{\Sigma}}}

\DeclareMathAlphabet{\mathsfit}{\encodingdefault}{\sfdefault}{m}{sl}
\SetMathAlphabet{\mathsfit}{bold}{\encodingdefault}{\sfdefault}{bx}{n}

\newcommand{\KL}{D_{\mathrm{KL}}}

\begin{document}
\doparttoc 
\faketableofcontents 

\title[Article Title]{MatterGen: a generative model for inorganic materials design}

\author[1]{\fnm{Claudio} \sur{Zeni}} \equalcont{}
\author[1]{\fnm{Robert} \sur{Pinsler}} \equalcont{}
\author[1]{\fnm{Daniel} \sur{Z{\"u}gner}} \equalcont{}
\author[1]{\fnm{Andrew} \sur{Fowler}} \equalcont{}
\author[1]{\fnm{Matthew} \sur{Horton}} \equalcont{}
\author[1]{\fnm{Xiang} \sur{Fu}}
\author[1]{\fnm{Aliaksandra} \sur{Shysheya}}
\author[1]{\fnm{Jonathan} \sur{Crabb{\'e}}}
\author[1]{\fnm{Lixin} \sur{Sun}}
\author[1]{\fnm{Jake} \sur{Smith}}
\author[1]{\fnm{Bichlien} \sur{Nguyen}}
\author[1]{\fnm{Hannes} \sur{Schulz}}
\author[1]{\fnm{Sarah} \sur{Lewis}}
\author[1]{\fnm{Chin-Wei} \sur{Huang}}
\author[1]{\fnm{Ziheng} \sur{Lu}}
\author[1]{\fnm{Yichi} \sur{Zhou}}
\author[1]{\fnm{Han} \sur{Yang}}
\author[1]{\fnm{Hongxia} \sur{Hao}}
\author[1]{\fnm{Jielan} \sur{Li}}
\author*[1]{\fnm{Ryota} \sur{Tomioka} \email{ryoto@microsoft.com}}\equalcont{}
\author*[1]{\fnm{Tian} \sur{Xie} \email{tianxie@microsoft.com}}\equalcont{Equal contribution; non-corresponding authors are listed in random order.}

\affil[1]{\orgdiv{Microsoft Research AI4Science}}

\abstract{
The design of functional materials with desired properties is essential in driving technological advances in areas like energy storage, catalysis, and carbon capture \cite{zhao2020designing,zhao2019theory,osman2021recent}. Generative models provide a new paradigm for materials design by directly generating entirely novel materials given desired property constraints. 
Despite recent progress, current generative models have low success rate in proposing stable crystals, or can only satisfy a very limited set of property constraints \cite{cdvae,zhao2023physics,kim2020generative,long2021constrained,zheng2023towards,yang2023scalable,noh2019inverse,court20203,antunes2023crystal,ai4science2023crystal}. Here, we present \mbox{MatterGen}, a model that generates stable, diverse inorganic materials across the periodic table and can further be fine-tuned to steer the generation towards a broad range of property constraints. To enable this, we introduce a new diffusion-based generative process that produces crystalline structures by gradually refining atom types, coordinates, and the periodic lattice. 
We further introduce adapter modules to enable fine-tuning towards any given property constraints with a labeled dataset. Compared to prior generative models, structures produced by \mbox{MatterGen} are more than twice as likely to be novel and stable, and more than 15 times closer to the local energy minimum. After fine-tuning, \mbox{MatterGen} successfully generates stable, novel materials with desired chemistry, symmetry, as well as mechanical, electronic and magnetic properties. 
Finally, we demonstrate multi-property materials design capabilities by proposing structures that have both high magnetic density and a chemical composition with low supply-chain risk. We believe that the quality of generated materials and the breadth of \mbox{MatterGen}'s capabilities represent a major advancement towards creating a universal generative model for materials design. 
}

\maketitle

\section{Introduction} \label{sec1}

The rate at which we can discover better materials has a major impact on the pace of technological innovation in areas such as carbon capture, semiconductor design, and energy storage.
Traditionally, most novel materials have been found through experimentation and human intuition, which require long iteration cycles and are limited by the number of candidates that can be tested.
Thanks to the advance of high throughput screening \cite{curtarolo2013high}, open material databases \cite{jain2013commentary,curtarolo2012aflow,kirklin2015open,choudhary2020joint,talirz2020materials}, machine-learning-based property predictors \cite{xie2018crystal,chen2019graph}, and \glspl{MLFF} \cite{unke2021machine,chen2022universal}, it has become increasingly common to screen hundreds of thousands of materials to identify promising candidates \cite{jun2022lithium,rosen2022high,zhong2020accelerated,merchant2023scaling}.
However, screening-based methods are still fundamentally limited by the number of known materials. 
The largest explorations of previously unknown crystalline materials are in the orders of $10^6$--$10^7$ materials \cite{shen2022reflections,chen2022universal,schmidt2022large,merchant2023scaling}, which is only a tiny fraction of the number of potential stable inorganic compounds ($10^{10}$ quaternary compounds only considering stoichiometry \cite{davies2016computational}).
Moreover, these methods cannot be efficiently steered towards finding materials with target properties.

Given these limitations, there has been an enormous interest in the inverse design of materials \cite{zunger2018inverse,sanchez2018inverse,schmidt2019recent,weiss2023guided}. The aim of inverse design is to directly generate material structures that satisfy possibly rare or even conflicting target property constraints, e.g., via generative models \cite{cdvae,yang2023scalable,ai4science2023crystal}, evolutionary algorithms \cite{allahyari2020coevolutionary}, and reinforcement learning \cite{law2022upper}. Generative models are particularly promising since they have the potential to efficiently explore entirely new structures, yet they can also be flexibly adapted to different downstream tasks. 
Despite recent progress, current generative models often fall short of producing stable materials according to \gls{DFT} calculations \cite{nouira2018crystalgan,ren2022invertible,cdvae,zhao2023physics}, are constrained by a narrow subset of elements \cite{noh2019inverse,court20203,zheng2023towards}, and/or can only optimize a very limited set of properties, mainly formation energy \cite{ren2022invertible,cdvae,zhao2023physics,yang2023scalable,ai4science2023crystal,sultanov2023data, lyngby2022data}.
 
In this study, we present \mbox{MatterGen}, a diffusion-based generative model for designing stable inorganic materials across the periodic table. 
MatterGen can further be fine-tuned via adapter modules to steer the generation towards materials with desired chemical composition, symmetry, and scalar property (e.g., band gap, bulk modulus, magnetic density) constraints.
Compared to the previous state-of-the-art generative model for materials \cite{cdvae}, \mbox{MatterGen} more than doubles the percentage of generated \gls{SUN} materials, and generates structures that are more than 15 times closer to their ground-truth structures at the \gls{DFT} local energy minimum.
When fine-tuned, \mbox{MatterGen} often generates more \gls{SUN} materials in target chemical systems than well-established methods like substitution and \gls{RSS}, is capable of generating highly symmetric structures given the desired space groups, and directly generates \gls{SUN} materials that satisfy target mechanical, electronic, and magnetic property constraints.
Finally, we showcase \mbox{MatterGen}'s ability to design materials given multiple property constraints by generating promising materials that have both high magnetic density and a chemical composition with low supply-chain risk.

    \section{Results}\label{sec:results}

\begin{figure}[t]
    \centering
    \includegraphics[width=\linewidth]{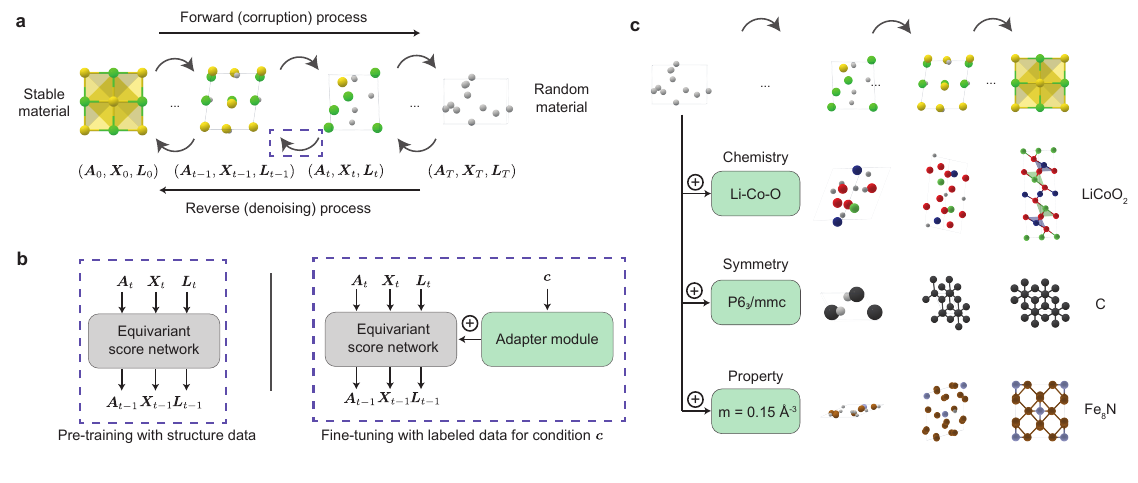}
    \caption{\textbf{Inorganic materials design with \mbox{MatterGen}.} \textbf{(a)} \mbox{MatterGen} generates stable materials by reversing a corruption process through iteratively denoising an initially random structure. The forward diffusion process is designed to independently corrupt atom types $\mA$, coordinates $\mX$, and the lattice $\mL$ to approach a physically motivated distribution of random materials. \textbf{(b)} An equivariant score network is pre-trained on a large dataset of stable material structures to jointly denoise atom types, coordinates, and the lattice. The score network is then fine-tuned with a labeled dataset through an adapter module that alters the model using the encoded property $\bm{c}$. \textbf{(c)} \mbox{MatterGen} can be fine-tuned to steer the generation towards materials with desired chemistry, symmetry, and scalar property constraints.}
    \label{fig:illustrative}
\end{figure}

\subsection{Diffusion process for crystalline material generation} \label{sec:mattergen}

In MatterGen, we introduce a novel diffusion process tailored for crystalline materials (\cref{fig:illustrative}(a)). Diffusion models generate samples by learning a score network to reverse a fixed corruption process \cite{dsm,ddpm,score_sde}. 
Corruption processes for images typically add Gaussian noise but crystalline materials have unique periodic structures and symmetries which demand a customized diffusion process. A crystalline material can be defined by its repeating unit, i.e., its unit cell, which encodes the atom types $\mA$ (i.e., chemical elements), coordinates $\mX$, and periodic lattice $\mL$ (\cref{app:representation,app:materials_symmetries}). We define a corruption process for each component that suits their own geometry and has physically motivated limiting noise distributions. More concretely, the coordinate diffusion respects the periodic boundary by employing a wrapped Normal distribution and approaches a uniform distribution at the noisy limit (\cref{app:coordinate-diffusion}). The lattice diffusion takes a symmetric form and approaches a distribution whose mean is a cubic lattice with average atomic density from the training data (\cref{app:lattice-diffusion}). The atom diffusion is defined in categorical space where individual atoms are corrupted into a masked state (\cref{app:atom-type-diffusion}). Given the corrupted structure, we learn a score network that outputs equivariant scores for atom type, coordinate, and lattice, respectively, which removes the need to learn the symmetries from data (\cref{app:architecture}). We refer to this network as the base model.

To generate materials with desired property constraints, we introduce adapter modules that can be used for fine-tuning the base model on an additional dataset with property labels (\cref{fig:illustrative}(b), more details in \cref{app:fine-tuning}). Fine-tuning is particularly appealing as it still works well if the labeled dataset is small compared to unlabeled structure datasets, which is often the case due to the high computational cost of calculating properties.
The adapter modules are tunable components injected into each layer of the base model to alter its output depending on the given property label \cite{zhang2023adding}.
The resulting fine-tuned model is used in combination with classifier-free guidance \cite{ho2022classifier} to steer the generation towards target property constraints.
We apply this approach to multiple types of properties, producing a set of fine-tuned models that can generate materials with target chemical composition, symmetry, or scalar properties such as magnetic density (\cref{fig:illustrative}(c)).

\subsection{Generating stable, diverse materials}\label{sec:generating_stable_diverse}

We formulate learning a generative model for inverse materials design as a two-step process, where we first pre-train a general base model for generating stable, diverse crystals across the periodic table, and then we fine-tune this base model towards different downstream tasks.
In this section, we focus on the ability of \mbox{MatterGen}'s base model to generate stable, diverse materials, which we argue is a prerequisite for addressing any inverse materials design task. Since diversity is difficult to measure directly, we resort to quantifying \mbox{MatterGen}'s ability to generate \gls{SUN} materials, and provide an additional analysis of the quality and diversity of generated structures.
To train the base model, we curate a large, diverse dataset comprising 607,684 stable structures with up to 20 atoms recomputed from the \gls{MP} \cite{jain2013commentary} and \gls{Alex} \cite{schmidt2022dataset,schmidt2022large} datasets, which we refer to as \gls{Alex-MP-20}.
We consider a structure to be stable if its energy per atom after relaxation via \gls{DFT} is below the 0.1~eV/atom threshold of a reference dataset comprising 1,081,850 unique structures recomputed from the \gls{MP}, \gls{Alex}, and \gls{ICSD} datasets. We refer to this dataset as \gls{Alex-MP-ICSD}.
We consider a structure to be novel if it is not contained in \gls{Alex-MP-ICSD}. We adopt these definitions throughout unless stated otherwise. More details are in \cref{app:data,sec:rmsd-stability-uniqueness-novelty}.

\begin{figure}[tbh]
    \centering
    \includegraphics[width=\linewidth]{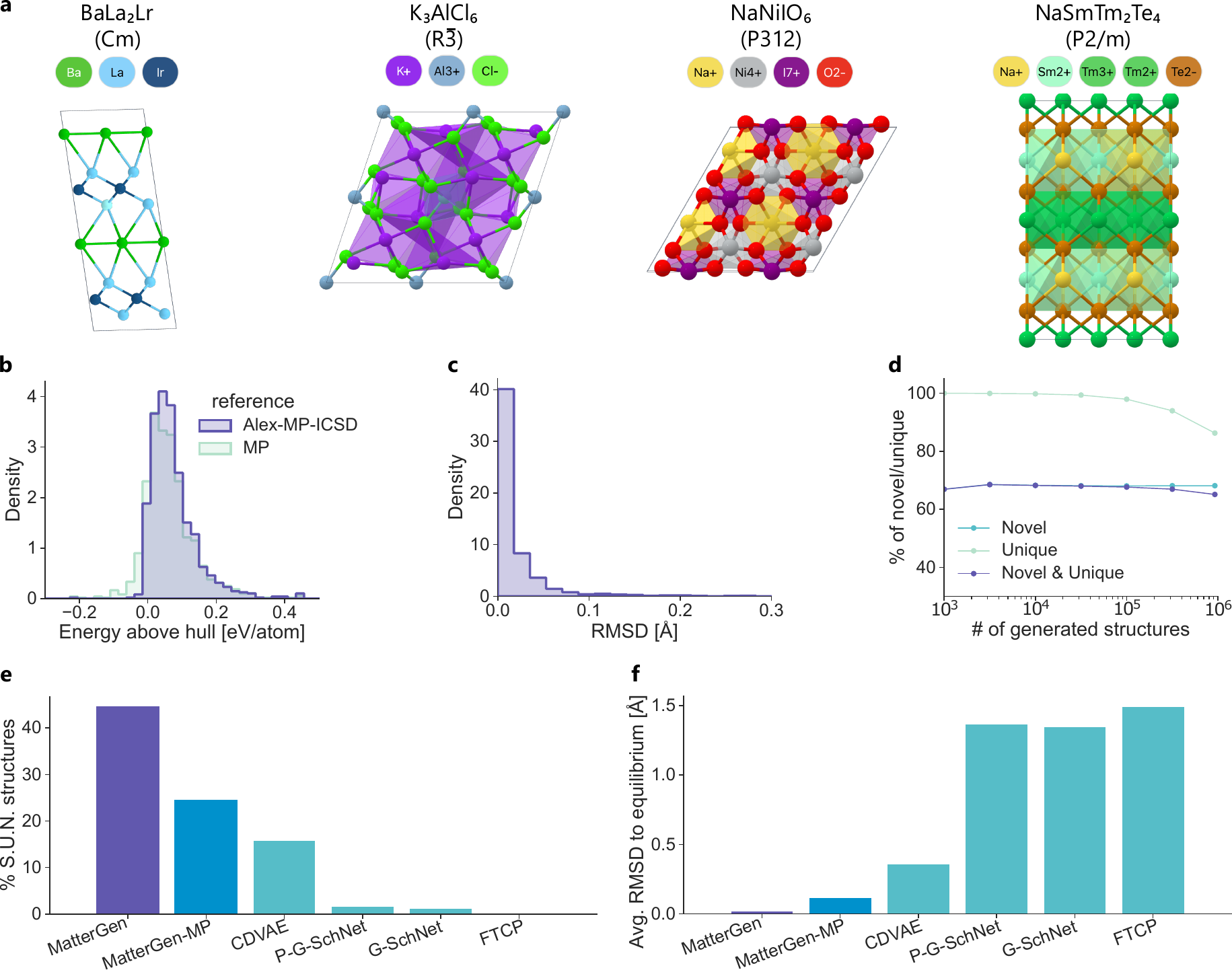}
    \caption{\textbf{Generating stable, unique and novel inorganic materials.} \textbf{(a)} Visualization of four randomly selected crystals generated by \mbox{MatterGen}, with corresponding chemical formula and space group symbols.
    \textbf{(b)} Distribution of energy above the hull using \gls{MP} and \gls{Alex-MP-ICSD} dataset as energy references, respectively. \textbf{(c)} Distribution of \gls{RMSD} between initial generated structures and \gls{DFT} relaxed structures. \textbf{(d)} Percentage of unique, novel structures as a function of number of generated structures. Novelty is defined with respect to \gls{Alex-MP-ICSD}. 
    \textbf{(e-f)} Percentage of \gls{SUN} structures (e) and average \gls{RMSD} between initial and \gls{DFT}-relaxed structures (f) for \mbox{MatterGen}, \mbox{MatterGen}-MP, and several baseline models, including CDVAE \cite{cdvae}, P-G-SchNet, G-SchNet \cite{gebauer2019symmetry}, and FTCP \cite{ren2022invertible}.}
    \label{fig:unconditional}
\end{figure}

\cref{fig:unconditional}(a) shows several random samples generated by \mbox{MatterGen}, featuring typical coordination environments of inorganic materials; see \cref{sec:structure_analysis_unconditional} for a more detailed analysis.
To assess stability, we perform \gls{DFT} calculations on 1024 generated structures.
\cref{fig:unconditional}(b) shows that 78~\% of generated structures fall below the 0.1~eV/atom threshold (13~\% below 0.0~eV/atom) of \gls{MP}'s convex hull, while 75~\%  fall below the 0.1~eV/atom threshold (3~\% below 0.0~eV/atom) of the combined \gls{Alex-MP-ICSD} hull (\cref{fig:unconditional}(b)). Further, 95~\% of generated structures have an \gls{RMSD} w.r.t.\ their \gls{DFT}-relaxed structures that is below \SI{0.076}{\angstrom} (\cref{fig:unconditional}(c)), which is almost one order of magnitude smaller than the atomic radius of the hydrogen atom (\SI{0.53}{\angstrom}). These results indicate that the majority of structures generated by \mbox{MatterGen} are stable, and very close to the \gls{DFT} local energy minimum.
We further investigate whether \mbox{MatterGen} can generate a substantial amount of unique and novel materials. We showcase in \cref{fig:unconditional}(d) that the percentage of unique structures is 100~\% when generating 1000 structures and only drops to 86~\% after generating one million structures, while novelty remains stable around 68~\%. 
This suggests that \mbox{MatterGen} is able to generate diverse structures without significant saturation even at a large scale, and that the majority of those structures are novel with respect to \gls{Alex-MP-ICSD}.

Moreover, we benchmark \mbox{MatterGen} against previous generative models for materials and show a significant improvement in performance. We focus on two key metrics: (1) the percentage of \gls{SUN} materials among generated samples, measuring the overall success rate of generating promising candidates, and (2) the average \gls{RMSD} between generated samples and their \gls{DFT}-relaxed structures, measuring the distance to equilibrium. 
We also compare to \mbox{MatterGen}-MP, which is a \mbox{MatterGen} model trained only on \gls{MP-20}, i.e., the same, smaller, dataset used by the other baselines. 
In \cref{fig:unconditional}(e-f), \mbox{MatterGen}-MP shows a 1.8 times increase in the percentage of \gls{SUN} structures and a 3.1 times decrease in average \gls{RMSD} compared with the previous state-of-the-art CDVAE \cite{cdvae}. 
When comparing \mbox{MatterGen} with \mbox{MatterGen}-MP, we observe a further 1.6 times increase in the percentage of \gls{SUN} structures and a 5.5 times decrease in \gls{RMSD} as a result of scaling up the training dataset.

In summary, we have shown that \mbox{MatterGen} is able to generate \gls{SUN} materials at a substantially higher rate compared to previous generative models while the generated structures are orders of magnitudes closer to their local energy minimum. Next, we fine-tune the pre-trained base model of \mbox{MatterGen} towards different downstream applications, including target chemistry (\cref{sec:chemical-system}), symmetry (\cref{sec:designing_space_group}), and scalar property constraints (\cref{sec:single-prop,sec:multi-prop}).

\subsection{Generating materials with target chemistry} \label{sec:chemical-system}
\begin{figure}
    \centering
    \includegraphics[width=1\linewidth]{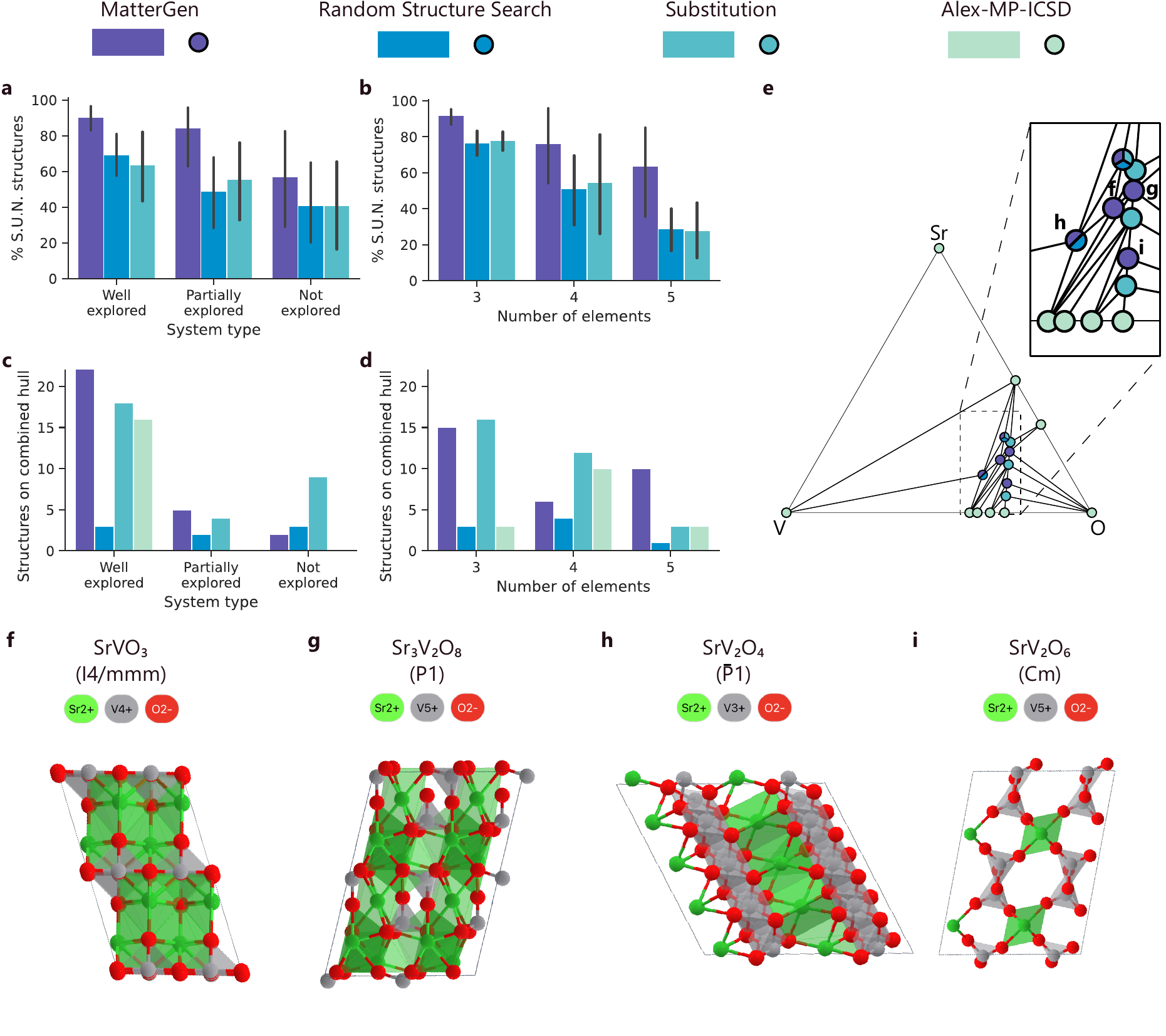}
    \caption{\textbf{Generating materials in target chemical system.} \textbf{(a-b)} Mean percentage of \gls{SUN} structures generated by \mbox{MatterGen} and baselines for 27 chemical systems, grouped by system type (a) and number of elements (b). Vertical black lines indicate maximum and minimum values.
    \textbf{(c-d)} Number of structures on the combined convex hull found by each method and in the \gls{Alex-MP-ICSD} dataset, grouped by system type (c) and number of elements (d).
    \textbf{(e)} Convex hull diagram for V-Sr-O, a well-explored ternary system.
    The dots represent structures on the hull, their coordinates represent the element ratio of their composition, and their color indicates by which method they were discovered.
    \textbf{(f-i)} Four of the five structures \mbox{MatterGen} discovered on the V-Sr-O hull depicted in (e), along with their composition and space group. 
    }
    \label{fig:chemsys}
\end{figure}
Finding the most stable material structures in a target chemical system (e.g., \mbox{Li-Co-O}) is crucial to define the true convex hull required for assessing stability, and indeed is one of the major challenges in materials design \cite{oganov2019structure}. 
The most comprehensive approach for this task is \textit{ab initio} \gls{RSS} \cite{pickard2011ab}, which has been used to discover many novel materials that were later experimentally synthesized \cite{oganov2019structure}.
The biggest drawback of \gls{RSS} is its computational cost, as the thorough exploration of even a ternary compound can require hundreds of thousands of \gls{DFT} relaxations.
In recent years, the combination of generating structures via \gls{RSS}, substitution or evolutionary methods with \glspl{MLFF} has proven successful in exploring chemical systems \cite{chen2022universal,merchant2023scaling,pickard2022ephemeral,conway2023search}.
Here, we evaluate the ability of \mbox{MatterGen} to explore target chemical systems by comparing it with substitution 
 \cite{chen2022universal} and \gls{RSS} \cite{pickard2011ab,zhu2021accelerating}.
We equip all methods with the \mbox{MatterSim}\cite{mattersim} \gls{MLFF}, which is used to pre-relax and filter the generated structures by their predicted stability before running more expensive \gls{DFT} calculations.
We fine-tune the \mbox{MatterGen} base model (\cref{sec:fine_tuning}) and steer the generated structures towards different target chemical systems and an energy above hull of 0.0~eV/atom.
We perform the benchmark evaluation for nine ternary, nine quaternary, and nine quinary chemical systems.
For each of these three groups, we pick three chemical systems at random from the following categories: well explored, partially explored, and not explored. See \cref{sec:supp_chem_sys} for additional details.
In \cref{fig:chemsys}(a-b) we see that \mbox{MatterGen} generates the highest percentage of \gls{SUN} structures for every system type and every chemical complexity.
As highlighted in \cref{fig:chemsys}(c), \mbox{MatterGen} also finds the highest number of unique structures on the combined convex hull in (1) `partially explored' systems, where existing known structures near the hull were provided during training, and in (2) `well-explored systems', where structures near the hull are known but were not provided in training.
While substitution offers a comparable or more efficient way to generate structures on the hull for ternary and quaternary systems, \mbox{MatterGen} achieves better performance on quinary systems, as shown in \cref{fig:chemsys}(d).
Remarkably, the strong performance of \mbox{MatterGen} in quinary systems was achieved with only 10,240 generated samples, compared to $\sim$70,000 samples for substitution and 600,000 for \gls{RSS}. This underscores the enormous efficiency gains that can be realized with generative models by proposing better initial candidates.
Finally, in \cref{fig:chemsys}(e) we show that \mbox{MatterGen} finds five novel structures on the combined hull for V-Sr-O---an example of a well-explored ternary system---while substitution finds four, and \gls{RSS} only two.
A few of the structures discovered by \mbox{MatterGen} are shown in \cref{fig:chemsys}(f-i), and are analyzed in-depth in \cref{subsec:structure_analysis_chemical_system}.

\subsection{Designing materials with target symmetry}\label{sec:designing_space_group}
\begin{figure}[tbh]
    \centering
\includegraphics[width=1\linewidth]{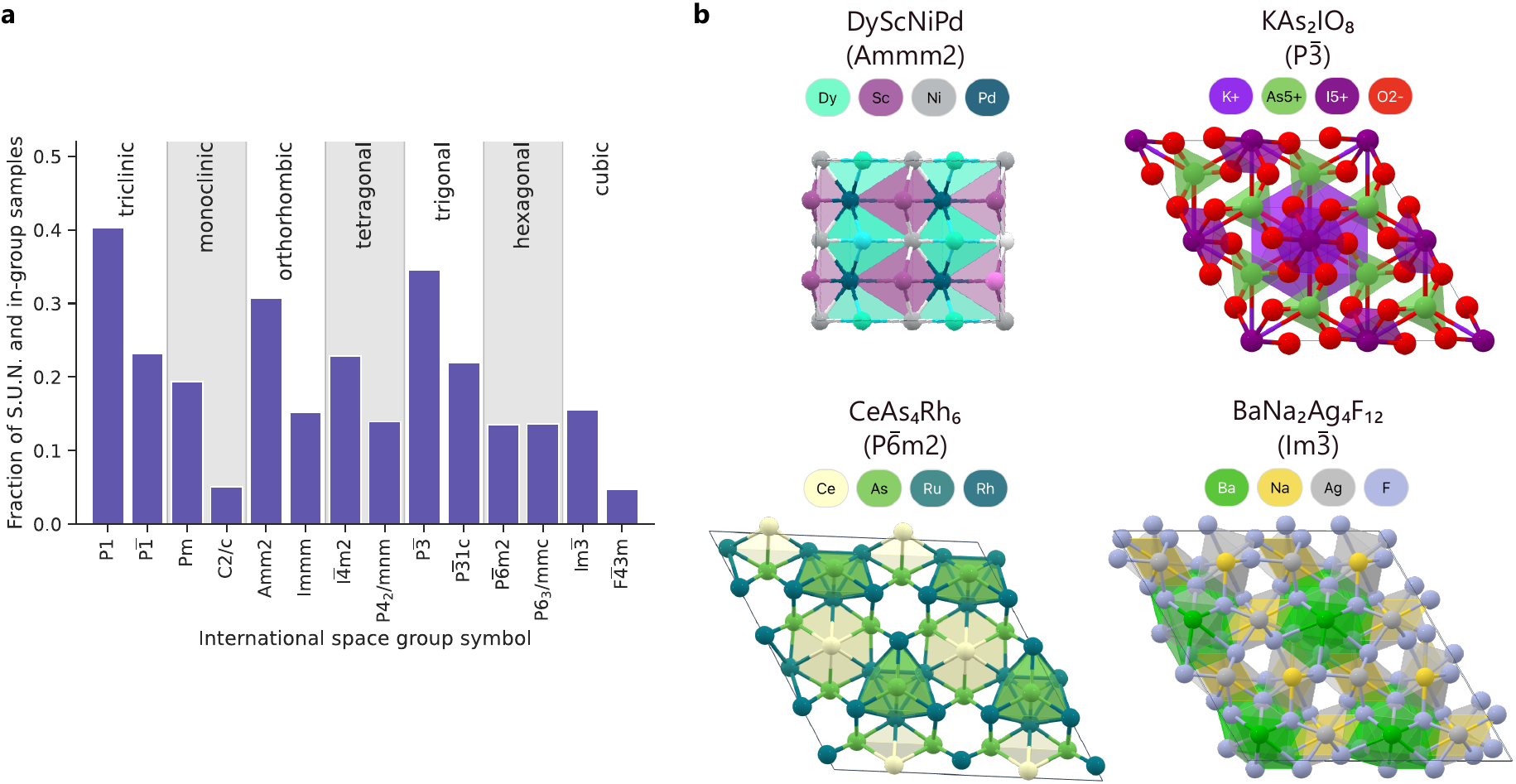}
    \caption{\textbf{Generating materials with target symmetry.}
    \textbf{(a)} Fraction of generated \gls{SUN} structures that belong to the target space group for 14 randomly chosen space groups spanning the seven lattice types.
    \textbf{(b)} Four randomly selected \gls{SUN} structures generated by \mbox{MatterGen}, along with their chemical formula and space group.
     }
    \label{fig:symmetry}
\end{figure}

The symmetry of a material directly affects its electronic and vibrational properties, and is a determining factor for its topological \cite{tang2019comprehensive} and ferroelectric \cite{smidt2020automatically} characteristics. 
The generation of \gls{SUN} materials with a given symmetry is a challenging task, as the symmetric arrangement of atoms in space is hard to enforce without resorting to explicit constraints based on already known materials.
Here, we assess \mbox{MatterGen's} ability to generate \gls{SUN} materials with a target symmetry by fine-tuning it on space group labels.
We choose 14 space groups at random from the subset of space groups that had at least 1000 entries in the training dataset, two for each of the seven crystal systems, and generate 256 structures per space group.
The results are shown in \cref{fig:symmetry}(a). On average, the fraction of generated \gls{SUN} structures that belong to the target space group is 20\%, and still surpassing $10\%$ for some of the most highly symmetric space groups that were chosen, e.g., $\text{P6}_3/\text{mmc}$ and $\text{Im}\bar{3}$. This is a notable result given that most previous generative models struggled in generating highly symmetric crystals \cite{cdvae,jiao2023crystal}.
In \cref{fig:symmetry}(b), we show four randomly generated \gls{SUN} structures from different space groups.
Additional details and results are provided in \cref{sec:supp_designing_space_group}.

\subsection{Designing materials with target magnetic, electronic, and mechanical properties}
\label{sec:single-prop}
\begin{figure}
    \centering
    \includegraphics[width=\linewidth]{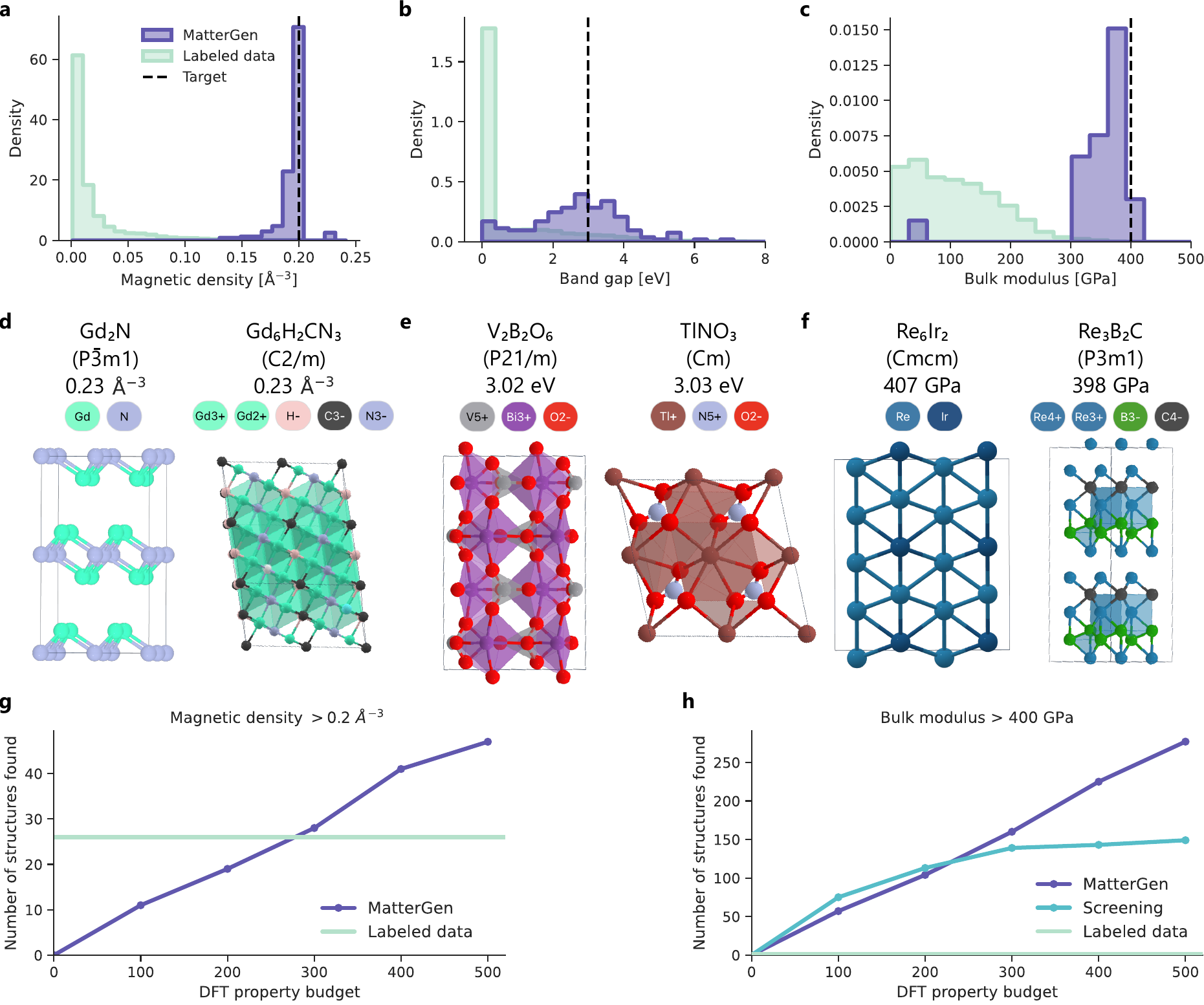}
    \caption{\textbf{Generating materials with target magnetic, electronic, and mechanical properties.} \textbf{(a-c)} Density of property values among (1) generated \gls{SUN} samples by \mbox{MatterGen}, and (2) structures in the labeled fine-tuning dataset for a magnetic, electronic, and mechanical property, respectively. The property target for \mbox{MatterGen} is shown as a black dashed line. Magnetic density values $< 10^{-3}~\text{\AA}^{-3}$ in (a) are excluded from the labeled data to improve readability. \textbf{(d-f)} Visualization of \gls{SUN} structures with the best property values generated by \mbox{MatterGen} for magnetic density (d), band gap (e), and bulk modulus (f). Alongside each structure, the chemical formula, space group and property value is shown.
    \textbf{(g-h)}
    Number of \gls{SUN} structures that satisfy target constraints found \mbox{MatterGen} compared to number of structures found by baselines across a range of \gls{DFT} property calculation budgets.}
    \label{fig:single-prop}
\end{figure}
There is an enormous need for new materials with improved properties across a wide range of real-world applications, e.g., for designing carbon capture technologies, solar cells, or semiconductors \cite{jun2022lithium,rosen2022high,zhong2020accelerated}. The classical screening-based approach starts from a set of candidates and selects the ones with the best properties. However, screening methods are unable to explore structures beyond the set of known materials.
Here, we demonstrate \mbox{MatterGen}'s ability to directly generate \gls{SUN} materials with target constraints on three different single-property inverse design tasks. These feature a diverse set of properties---magnetic, electronic, and mechanical---with varying degrees of available labeled data for fine-tuning the model.
In the first task, we aim to generate materials with high magnetic density, a prerequisite for permanent magnets. We fine-tune the model on 605,000 structures with \gls{DFT} magnetic density labels (calculated assuming ferromagnetic ordering) and then generate structures with a target magnetic density value of $0.20~\text{\AA}^{-3}$.
Second, we search for materials with a specific electronic property. We fine-tune the model on 42,000 structures with \gls{DFT} band gap labels, then sample materials with a target calculated band gap value of 3.0~eV.
Finally, we target structures with high bulk modulus---an important property for superhard materials. We fine-tune the model on only 5,000 labeled structures, and sample with a target value of 400 GPa.
While the tasks above were chosen to evaluate the generality of the model, we note that additional follow-up investigations would be required to assess the suitability of these materials for specific applications, e.g., a superhard material needs to satisfy additional constraints such as a high shear modulus, or a permanent magnet needs a suitable magnetic order and critical temperature.
See \cref{sec:supp-single-prop} for more details.

In \cref{fig:single-prop}(a-c), we highlight the significant shift in the distribution of property values among \gls{SUN} samples generated by \mbox{MatterGen} towards the desired targets, even when the targets are at the tail of the data distribution. In particular, this still holds true for properties where the number of \gls{DFT} labels available for fine-tuning the model is substantially smaller than the size of the unlabeled training data. In \cref{fig:single-prop}(d-f) we showcase the \gls{SUN} structures with the best property values generated by \mbox{MatterGen} for each task. See \cref{sec:supp-single-prop-analysis} for additional analysis.

Moreover, we assess how many \gls{SUN} structures satisfying extreme property constraints can be found by \mbox{MatterGen} when given a limited budget for \gls{DFT} property calculations. As a baseline, we count the number of materials in the labeled fine-tuning dataset that satisfy the constraint.
We also compare with a screening approach, which scans previously unlabeled materials for promising candidates. 
In contrast to the previous experiment, we fine-tune \mbox{MatterGen} with labels predicted by a machine learning property predictor---the same used for the screening baseline---when the dataset is not fully labeled.
As shown in \cref{fig:single-prop}(g), \mbox{MatterGen} is able to find up to 47 \gls{SUN} structures with magnetic density above $0.2~\text{\AA}^{-3}$, much more than the $26$ materials with such high property values in the fine-tuning dataset. Since the dataset is fully labeled, there is no screening baseline available. In \cref{fig:single-prop}(h), we see that \mbox{MatterGen} finds substantially more \gls{SUN} materials with high bulk modulus than screening. While the number of structures found by screening saturates with increasing budget, \mbox{MatterGen} keeps discovering \gls{SUN} structures at an almost constant rate. Given a budget of 500 \gls{DFT} property calculations, we find 277 \gls{SUN} structures (with 126 distinct compositions), almost double the number found with a screening approach (149, 79 distinct compositions).
In contrast, there are only two materials in the labeled fine-tuning dataset with such high bulk modulus values. Note that both \mbox{MatterGen} and screening produce multiple structures per composition that are unique according to our definition (\cref{sec:rmsd-stability-uniqueness-novelty}) but could potentially be alloys or solid solutions \cite{leeman2024challenges}.

\subsection{Designing low-supply-chain-risk magnets} \label{sec:multi-prop}

Most materials design problems require finding structures satisfying multiple property constraints. \mbox{MatterGen} can be fine-tuned to generate materials given any combination of constraints. Here, we showcase its ability to tackle materials design problems with multiple constraints by searching for low-supply-chain-risk magnets.
Since many existing high-performing permanent magnets contain rare earth elements that are subject to supply chain risks, there has been increasing interest in discovering rare-earth-free permanent magnets \cite{cui2018current}. We simplify the problem of finding such a magnet to finding materials with high magnetic density and a low \gls{HHI}. According to the U.S. Department of Justice and the Federal Trade Commission, a material with an \gls{HHI} score below 1500 is considered to have low supply chain risk \cite{gaultois2013data}.
Thus, we ask the model to generate materials with a magnetic density of $0.2~\si{\angstrom}^{-3}$ and an \gls{HHI} score of 1250. 

\begin{figure}[tbh]
    \centering
    \includegraphics[width=1\linewidth]{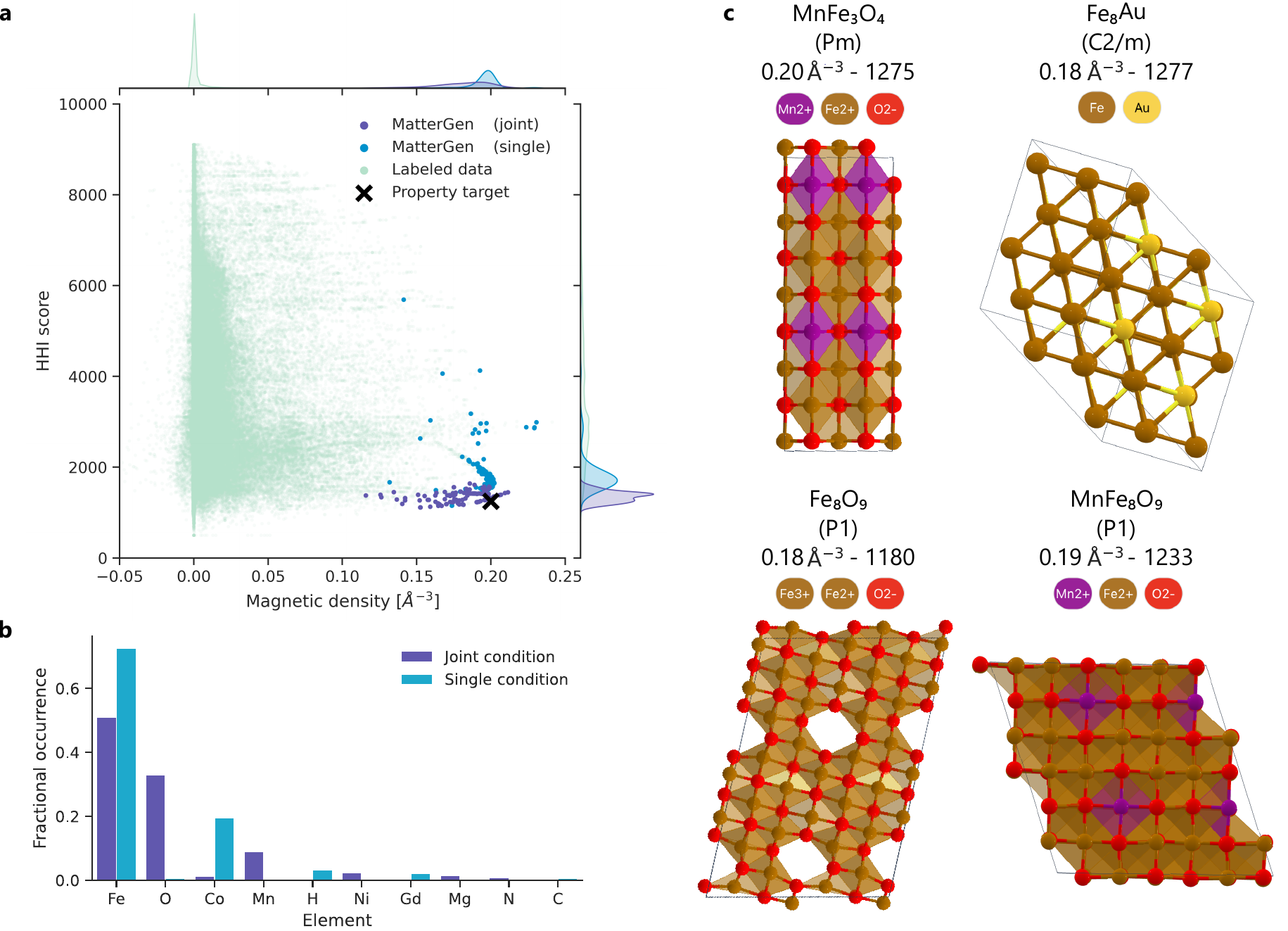}
    \caption{\textbf{Designing low-supply-chain-risk magnets.} \textbf{(a)} Distribution of \gls{SUN} structures generated by \mbox{MatterGen} when fine-tuned on the \gls{HHI} score (single) and on both \gls{HHI} score and magnetic density (joint), as well as structures from the labeled fine-tuning dataset. The property target of \mbox{MatterGen} is denoted as a black cross.
    \textbf{(b)} Occurrence of most frequent elements in \gls{SUN} structures for the two fine-tuned \mbox{MatterGen} models.
    \textbf{(c)} \gls{SUN} structures on the Pareto front for the joint property optimization task, along with their chemical composition, space group, magnetic density, and HHI score.}
    \label{fig:multi-prop}
\end{figure}

In \cref{fig:multi-prop}(a), we observe that \mbox{MatterGen} generates \gls{SUN} structures that are narrowly distributed around the target values, despite the labeled fine-tuning data being extremely scarce in that region. Compared to a model that only targets high magnetic density values (single), targeting both properties (joint) shifts the distribution of \gls{HHI} scores closer towards the desired target value while retaining high magnetic density values. 
\cref{fig:multi-prop}(b) showcases the effect of jointly targeting both properties on the distribution of elements found in the generated materials. Due to the lower \gls{HHI} scores, elements commonly employed for high-magnetic density materials that have supply chain issues, e.g., Cobalt (Co) and Gadolinium (Gd), are practically absent in the jointly generated structures. In contrast, these elements are still present in structures generated by the model only targeting materials with high magnetic density (single).

\section{Discussion}

Generative models are particularly promising for tackling inverse design tasks as they can potentially explore entirely \emph{novel} structures with desired properties in an efficient way.
However, generating the 3D structure of stable crystalline materials is challenging due to their periodicity and the interplay between atom types, coordinates, and lattice. 
MatterGen improves upon limitations of previous methods \cite{cdvae,jiao2023crystal} by introducing a joint diffusion process for atom types, coordinates, and lattice (\cref{sec:mattergen,app:atom-type-diffusion,app:coordinate-diffusion,app:lattice-diffusion}), which---combined with the substantially larger training dataset \gls{Alex-MP-20}---drastically increase the stability, uniqueness, and novelty of generated materials.
Thanks to the introduction of the adapter modules (\cref{sec:fine_tuning}), \mbox{MatterGen} can further be fine-tuned to generate \gls{SUN} structures satisfying target constraints across a wide range of properties, with performance improvements over widely-employed methods such as \gls{MLFF}-assisted \gls{RSS} and substitution (\cref{sec:chemical-system}), as well as ML-assisted screening (\cref{sec:single-prop}).

Despite these advances, \mbox{MatterGen} could still be improved in several ways.
For example, we observe that the model disproportionately generates structures with P1 symmetry compared to the training data, indicating a tendency for generating less symmetric structures, especially for larger crystals (see discussion in \cref{app:analysis}).
We hypothesize that further improvements on the denoising process, the backbone architecture, and the expansion of the training dataset could enable the model to overcome such issues. We also acknowledge that our extensive evaluations only cover some of the criteria required for real-world applications, with experimental validation and characterization being the ultimate test \cite{leeman2024challenges}. We discuss the challenges in evaluating the quality of crystalline materials from generative models in \cref{app:analysis}.

Overall, we believe that the breadth of \mbox{MatterGen}'s capabilities and the quality of generated materials represent a major advancement towards creating a universal generative model for materials with high real-world impact. Given the transformative effect of generative models in domains like image generation \cite{ramesh2022hierarchical} and protein design \cite{watson2023novo}, we envision that generative models like \mbox{MatterGen} will have a major impact in materials design in the coming years. As such, we are excited about the many directions in which \mbox{MatterGen} could be further extended.
For instance, \mbox{MatterGen} could be expanded to cover a broader class of materials ranging from catalyst surfaces to metal organic frameworks, enabling us to tackle challenging problems like nitrogen fixation \cite{guo2019electrochemical} and carbon capture \cite{sumida2012carbon}. The property constraints can be extended to non-scalar quantities like the band structure or X-ray diffraction (XRD) spectrum, which would further enable applications ranging from band engineering to the prediction of atomic structures of experimentally-measured XRD spectra of unknown samples.

\backmatter

\bmhead*{Supplementary information}
Additional explanations and details regarding the \mbox{MatterGen} architecture, fine-tuning approach, datasets, and results can be found in the supplementary information.

\bmhead*{Acknowledgments}
We are grateful for many insightful discussions with members from the Materials Project~\citep{jain2013commentary}, and to Chris Pickard for providing helpful feedback. We would also like to thank our colleagues from Microsoft Research AI4Science for helpful discussions and support, including Andrew Foong, Karin Strauss, Keqiang Yan, Cristian Bodnar, Rianne van den Berg, Frank No{\'e}, Marwin Segler, Elise van der Pol, and Max Welling. We are also grateful for useful feedback from the Microsoft Azure Quantum team, including Chi Chen, Leopold Talirz and Nathan Baker. Finally, we thank the AI on Xbox Team for providing part of the compute resources required for this work.

\section*{Declarations}

\paragraph*{Author contributions}

AF, MH, RP, RT, TX, CZ and DZ (alphabetically ordered) conceived the study, implemented the methods, performed experiments, and wrote the manuscript. XF led the development of the adapter modules. SS implemented and ran the symmetry conditioned generation. JS implemented the band gap workflow. BN proposed the task of low-supply-chain risk magnets. ZL, YZ, HY, HH, and JL developed the machine learning force field. XF, SS, JC, LS, JS, BN, HS, SL, C-WH, ZL, YZ, HY, HH, and JL helped with implementing the methods, conducting experiments, and writing the manuscript. TX and RT led the research.

\begin{appendices}
\appendix
\def\coords{\mX}  
\def\cartcoords{\tilde{\coords}}  
\def\coord{\vx}  
\def\cartcoord{\tilde{\coord}}  
\def\lattice{\mL}  
\def\lvec{\vl}  
\def\type{a}  
\def\typevec{\va}  
\def\types{\mA}  
\def\typespace{\mathbb{A}}  
\def\mat{\mM}  
\def\cat{\mathrm{Cat}}  
\def\tdisc{t}  
\def\nodesel{i}  
\def\othernodesel{j}  
\def\hiddendim{d}  
\def\numnodes{n}  
\newcommand{\tindex}[1]{_{#1}}  
\newcommand{\nindex}[1]{^{#1}}  
\newcommand{\edgeset}{\mathcal{E}}
\newcommand{\transpose}{^\top}

\part{Appendix} 
\parttoc 

\newpage

\section{Diffusion model for periodic materials} \label{app:diffusion}

This section contains additional model details, following the notation listed in \cref{tab:notation_table}.

\begin{table}[h!]
    \centering
    \begin{tabular}{l|l}
    \toprule
    \multicolumn{2}{c}{General notation} \\ \midrule
    $\numnodes \in \mathbb{N}$ & Number of atoms in a crystal \\
    $\mat = (\coords, \types, \lattice)$ & A crystal structure \\
    $\coords \in [0, 1)^{3 \times \numnodes}$ & Fractional atomic coordinates \\
    $\cartcoords \in \mathbb{R}^{3 \times \numnodes}$ & Cartesian atomic coordinates \\
    $\types \in \mathbb{A}^\numnodes$ & Atomic species in a crystal \\
    $\lattice = (\lvec\nindex{1}, \lvec\nindex{2}, \lvec\nindex{3}) \in \mathbb{R}^{3\times3}$ & The unit cell lattice matrix \\
    $\lvec\nindex{j} \in \mathbb{R}^3$, $j \in \{1, 2, 3\}$ & The $j$-th lattice vector \\
    $\mV = (\vv_1, \vv_2, \ldots \vv_N) \in \mathbb{R}^{d \times N}$  & Concatenation of $N$ $d$-dimensional column vectors into a matrix \\
    $\edgeset \subset \{1, 2, \ldots, \numnodes\}^2 \times \mathbb{Z}^3 $ & Set of edges in a material \\
    $\nodesel, \othernodesel \in \{1, 2, \ldots, \numnodes\}$ & Index of an atom in a material \\
    $\hiddendim \in \mathbb{N}$ & The number of hidden dimension in our GNN\\
    $\bm{1}_n \in\mathbb{R}^{n}$ & $n$-dimensional column vector containing ones \\
    \bottomrule
    \multicolumn{2}{c}{Diffusion notation} \\
    \midrule
    $\tdisc \in {1, 2, \ldots, T}$ & Diffusion timestep  \\
    $T \in \mathbb{N}$ & Number of time discretization steps for the diffusion process \\
    $q(\coord\tindex{0})$ & The data distribution \\
    $q(\coord\tindex{\tdisc} | \coord\tindex{\tdisc-1})$ & Single-step diffusion transition kernel \\
    $q(\coord\tindex{\tdisc} | \coord\tindex{0})$ & One-shot diffusion kernel \\
    $q(\coord\tindex{T})$ & Prior (noise) distribution \\
    $\vs_\vtheta(\cdot, \tdisc)$ & Score model \\
    $\vs_{\coords, \vtheta}(\cdot, \tdisc)$ & Score model for atomic coordinates\\
    $\log p_\vtheta(\types\tindex{0} | \coords\tindex{\tdisc}, \lattice\tindex{\tdisc}, \types\tindex{\tdisc}, \tdisc)$ & Predicted logits for atom types at $\tdisc=0$. \\
    $\vs_{\lattice, \vtheta}(\cdot, \tdisc)$ & Score model for lattice\\
    $\vz$ & Standard Gaussian noise $\vz \sim \mathcal{N}(\vzero, \mI)$ \\
    \bottomrule
    \end{tabular}
    \caption{Table of notations}
    \label{tab:notation_table}
\end{table}

\subsection{Representation of periodic materials} \label{app:representation}

Any crystal structure can be represented by some repeating unit (called the \emph{unit cell}) that tiles the entire 3D space. The unit cell itself contains a number of atoms that are arranged inside of it. Thus, we use the following universal representation for a material $\mat$:
\begin{equation}
    \mat = \left(\types, \coords, \lattice \right),
    \label{eq:crystaldef}
\end{equation}
where $\types = (\type\nindex{1}, \type\nindex{2}, \ldots, \type\nindex{\numnodes})\transpose \in \typespace^\numnodes$ are the atomic species of the atoms inside the unit cell; $\lattice = (\lvec\nindex{1}, \lvec\nindex{2}, \lvec\nindex{3}) \in \mathbb{R}^{3\times 3}$ is the lattice, i.e., the shape of the repeating unit cell; and $\coords = (\coord\nindex{1}, \coord\nindex{2}, \ldots, \coord\nindex{\numnodes}) \in [0, 1)^{3 \times \numnodes}$ are the \emph{fractional} coordinates of the atoms inside the unit cell. 

The lattice $\lattice$ is a parallelepiped defined by the three lattice vectors $\lvec\nindex{1}, \lvec\nindex{2},$ and $\lvec\nindex{3}$. It can thus be compactly represented as a single $3 \times 3$ matrix with the three lattice vectors as its columns. The volume of a lattice is given by $\mathrm{Vol}(\lattice) = |\det \lattice|$. Any physically sensible crystal must have a unit cell with nonzero volume, hence, we require any lattice matrix to be non-singular.

Fractional coordinates express the location of an atom using the lattice vectors as the basis vectors. For instance, an atom with fractional coordinates $\coord = (0.2, 0.3, 0.5)\transpose$ has Cartesian coordinates $\cartcoord= 0.2 \lvec\nindex{1} + 0.3 \lvec\nindex{2} + 0.5 \lvec\nindex{3}$.
The periodicity in fractional coordinates is defined by the (flat) unit hypertorus, i.e., we have the equivalence relation $\coord \sim \coord + \vk, \vk \in \mathbb{Z}^{3}$.
We can convert between fractional coordinates $\coords $ and Cartesian coordinates $\cartcoords$ as follows:
\begin{align}
    \cartcoords &= \lattice \coords \label{eq:frac_to_cart}, \\
    \coords &= \lattice^{-1} \cartcoords \label{eq:cart_to_frac}.
\end{align}

\subsection{Invariance and equivariance in periodic materials}
\label{app:materials_symmetries}
The energy per atom $\epsilon(\mat) = E(\mat)/n$ of a periodic material $\mat = (\coords, \lattice, \types)$ has several invariances.
%
\begin{itemize}
\item Permutation invariance: $\epsilon(\coords, \lattice, \types) = \epsilon(\mP(\coords), \lattice, \mP(\types))$ for every permutation matrix $\mP$.
\item Translation invariance: $\epsilon(\coords, \lattice, \types) = \epsilon(\coords + \vt, \lattice, \types)$ for every $\vt \in \mathbb{R}^3$.
\item Rotation invariance: $\epsilon(\coords, \lattice, \types) = \epsilon(\coords, \mR(\lattice), \types)$ for every rotation matrix $\mR \in O(3)$.
\item Periodic cell choice invariance: $\epsilon(\coords, \lattice, \types) = \epsilon(\mC^{-1}\coords, \lattice \mC , \types)$, where $\mC$ triangular with $\det \mC = 1$ and $\mC \in \mathbb{Z}^{3 \times 3}$. See \cref{fig:equivalent_lattices} for an example.
\item Supercell invariance: $\epsilon(\coords, \lattice, \types) = \epsilon \left( \bigoplus_{i=0}^{\det(\mC)}  \mC^{-1}(\coords + \vk_i \bm{1}_n\transpose), \lattice \mC , \bigoplus_{i=0}^{\det(\mC)} \types \right)$, where $\mC$ is a 3$\times$3 diagonal matrix with positive integers on the diagonal, $\vk_i \in \mathbb{N}^3$  indexes the cell repetitions in the three lattice components, and  $\bigoplus$ indicates concatenation.
\end{itemize}
Forces are instead equivariant to permutation and rotation, while being invariant to translation and periodic cell choice.
Stress tensors are similarly invariant to permutation, translation, supercell choice, and periodic cell choice; while being equivariant to rotation (see \cref{sec:lattice_scores} for additional details).

\subsection{Diffusion model background and notation }
\label{app:diffusion-model-background}
Diffusion models \citep{ddpm,sohl-dickstein15,dsm,d3pm} are a class of generative models that learn to revert a diffusion process. The diffusion process (also called the \emph{forward} process) gradually corrupts an input sample $\coord\tindex{0}$ via transition kernels $q(\coord\tindex{\tdisc } | \coord\tindex{\tdisc - 1})$\footnote{We follow the convention in machine learning literature that the functional forms of (conditional) probability density functions depend on the variables that appear as arguments. For example, $q(\coord\tindex{\tdisc } | \coord\tindex{\tdisc - 1})$ could be written as $q_{\coords\tindex{\tdisc}| \coords\tindex{\tdisc-1}}(\coord\tindex{\tdisc}| \coord\tindex{\tdisc-1})$ to make the dependence of the functional form on $t$ explicit, but we avoid this to prevent clutter.}, defining a Markov chain $\coord\tindex{0}\rightarrow\coord\tindex{1}\rightarrow \cdots\rightarrow \coord\tindex{T}$, where $T \in \mathbb{N}$ is the number of diffusion steps and $1 \leq \tdisc \leq T$. Here, we cover the typical case where the data is continuous-valued and the transition kernels are normal distributions. See \cref{app:atom-type-diffusion} for details on discrete diffusion models.

The transition kernels are of the general form $q(\coord\tindex{\tdisc} | \coord\tindex{\tdisc - 1}) = \mathcal{N}(f(\coord\tindex{\tdisc-1}, \tdisc), \sigma\tindex{\tdisc}^2 \mI)$, where $f(\coord\tindex{\tdisc-1}, \tdisc): \mathbb{R}^{d} \rightarrow \mathbb{R}^{d}$ is affine in $\coord\tindex{\tdisc-1}$. This implies that the one-shot transition kernel $q(\coord\tindex{\tdisc} | \coord\tindex{0})$ is also Gaussian, and for popular choices $f(\cdot, \tdisc)$ the mean and variance are known in closed form. This enables us to efficiently obtain a noisy sample $\coord\tindex{\tdisc}$ at an arbitrary time step $\tdisc$ during training.

Diffusion models are optimized to approximate the score function of the noise distributions $q(\coord\tindex{\tdisc} | \coord\tindex{0})$ for any $1 \leq \tdisc \leq T$:
\begin{equation} \label{eq:score_matching_objective}
    \vtheta^* = \underset{\vtheta}{\arg\min}\sum_{\tdisc=1}^T \sigma\tindex{\tdisc}^2 \mathbb{E}_{q(\coord\tindex{0})} \mathbb{E}_{q(\coord\tindex{\tdisc} | \coord\tindex{0})}\left[\|\vs_\vtheta(\coord\tindex{\tdisc}, \tdisc) - \nabla_{\coord\tindex{\tdisc}}\log q(\coord\tindex{\tdisc} | \coord\tindex{0}) \|_2^2 \right],
\end{equation}
where $\vs_\vtheta(\coord, \tdisc): \mathbb{R}^d \times \mathbb{R}_{+} \rightarrow \mathbb{R}^d$ is called the \emph{score model}. It is standard practice \citep{ddpm,dsm} to parameterize the model to predict the \emph{noise} $\epsilon\tindex{\tdisc}=-\sigma\tindex{\tdisc}\nabla_{\coord\tindex{\tdisc}}\log q(\coord\tindex{\tdisc} | \coord\tindex{0})$ instead of the score, since the magnitude of $\epsilon\tindex{\tdisc} \sim \mathcal{N}(\vzero, \mI)$ is independent of the diffusion time step $\tdisc$.

The forward diffusion process is designed such that $q(\coord\tindex{T} | \coord\tindex{0}) \approx q(\coord\tindex{T})$, where $q(\coord\tindex{T})$ is a prior distribution that is easy to sample from (e.g., Gaussian).

In this work we leverage two popular diffusion processes for continuous data, i.e., the variance-exploding diffusion  \citep{dsm,dsm_improved} and the variance-preserving diffusion \citep{ddpm,sohl-dickstein15} process, which we briefly explain in the following.

\paragraph*{Variance-exploding diffusion} This is the diffusion process used in \gls{DSM} \citep{dsm}. We define a sequence of exponentially increasing standard deviations $\sigma\tindex{\textnormal{min}} = \sigma\tindex{1}, \ldots, \sigma\tindex{T} = \sigma\tindex{\textnormal{max}}$ that define the transition kernels:
\begin{equation}
    q(\coord\tindex{\tdisc} | \coord\tindex{\tdisc - 1}) = \mathcal{N}\left(\coord\tindex{\tdisc}, \left(\sigma\tindex{\tdisc}^2 - \sigma\tindex{\tdisc - 1}^2\right) \mI \right), \qquad q(\coord\tindex{\tdisc} | \coord\tindex{0}) = \mathcal{N}\left(\coord\tindex{0}, \sigma\tindex{\tdisc}^2 \mI \right) .
\end{equation}
We can generate a sample using the learned model via annealed Langevin dynamics \citep{dsm,dsm_improved} or ancestral sampling from the graphical model $\prod_{\tdisc=1}^T p_\vtheta\left(\coord\tindex{\tdisc-1} | \coord\tindex{\tdisc} \right)$ \citep{score_sde}:
\begin{equation} \label{eq:dsm_reverse}
    \coord\tindex{\tdisc - 1} = \coord\tindex{\tdisc} + (\sigma\tindex{\tdisc}^2 - \sigma\tindex{\tdisc - 1}^2)\vs_{\vtheta^*}(\coord\tindex{\tdisc}, \tdisc) + \vz\sqrt{\sigma\tindex{\tdisc}^2 - \sigma\tindex{\tdisc - 1}^2},
\end{equation}
where $\coord\tindex{T} \sim \mathcal{N}\left(\vzero, \sigma\tindex{T}^2 \mI \right)$, and $\vz \sim \mathcal{N}\left(\vzero, \mI \right)$ is standard Gaussian noise. In \cref{app:coordinate-diffusion} we explain how we leverage variance-exploding diffusion in the diffusion process of the fractional coordinates.

\paragraph*{Variance-preserving diffusion} This is the diffusion process used to train \glspl{DDPM} \citep{ddpm,sohl-dickstein15}. In variance-preserving diffusion we define a sequence of positive noise scales $0 < \beta\tindex{1}, \beta\tindex{2}, \ldots \beta\tindex{T} < 1$ to obtain transition kernels of the form 
\begin{equation}
    q(\coord\tindex{\tdisc} | \coord\tindex{\tdisc-1}) = \mathcal{N}\left(\sqrt{1 - \beta\tindex{\tdisc}} \coord\tindex{\tdisc-1}, \beta\tindex{\tdisc} \mI \right), \qquad q(\coord\tindex{\tdisc} | \coord\tindex{0}) = \mathcal{N}\left(\sqrt{\bar{\alpha}\tindex{\tdisc}} \coord\tindex{0}, \left(1-\bar{\alpha}\tindex{\tdisc}\right)\mI \right),
\end{equation}
where $\bar{\alpha}\tindex{\tdisc} = \prod_{i=1}^{\tdisc}\left(1-\beta\tindex{\tdisc} \right)$. Sampling from a model trained to revert the variance-preserving diffusion process also works via \emph{ancestral sampling} from the graphical model $\prod_{\tdisc=1}^T p_\vtheta\left(\coord\tindex{\tdisc-1} | \coord\tindex{\tdisc} \right)$:
\begin{equation} \label{eq:ddpm_reverse}
    \coord\tindex{\tdisc-1} = \frac{1}{\sqrt{1 - \beta\tindex{\tdisc}}}\left(\coord\tindex{\tdisc} + \beta\tindex{\tdisc}\vs_{\vtheta^*}(\coord\tindex{\tdisc}, \tdisc) \right) + \sqrt{\beta\tindex{\tdisc}}  \vz,
\end{equation}
starting from $\coord\tindex{T} \sim \mathcal{N}\left(\vzero, \mI \right)$, where $\vz \sim \mathcal{N}\left(\vzero, \mI \right)$ is standard Gaussian noise. See \cref{app:lattice-diffusion} for details about how we leverage variance-preserving diffusion in the diffusion process of the lattice.

\subsection{Joint diffusion process} \label{app:joint-process}
\label{sec:joint_diffusion_details}
To apply the construction of a diffusion process described in \cref{app:diffusion-model-background} to crystal structures described in \cref{app:representation}, we define the forward process through a Markov chain $\mat\tindex{0}\rightarrow \mat\tindex{1}\rightarrow \cdots\rightarrow\mat\tindex{T}$ via a transition kernel that diffuses the atom coordinates, atom types, and the lattice independently as follows:
\begin{align}
    &q(\types\tindex{\tdisc+1},
    \coords\tindex{\tdisc+1}, \lattice\tindex{\tdisc+1} | 
    \types\tindex{\tdisc},
    \coords\tindex{\tdisc},
    \lattice\tindex{\tdisc}) \notag\\
    &\quad= q(\types\tindex{\tdisc+1} | \types\tindex{\tdisc})
    q(\coords\tindex{\tdisc+1} | \coords\tindex{\tdisc})  q(\lattice\tindex{\tdisc+1} | \lattice\tindex{\tdisc}) \qquad(t=0,1,\ldots,T-1). \label{eq:diffusion_factorization}
\end{align}

In addition, the noise distributions of  atom species $\types$ and the fractional coordinates $\coords$ factorize into the diffusion of the individual atoms:
\begin{equation*}
    q(\types\tindex{\tdisc+1} | \types\tindex{\tdisc}) = \prod_{\nodesel=1}^{\numnodes}q(\type^\nodesel\tindex{\tdisc+1} | \type^\nodesel\tindex{\tdisc}),\qquad q(\coords\tindex{\tdisc+1} | \coords\tindex{\tdisc}) = \prod_{\nodesel=1}^{\numnodes}q(\coord^\nodesel\tindex{\tdisc+1} | \coord^\nodesel\tindex{\tdisc}).
\end{equation*}
Note that the factorization of the forward diffusion process does not imply that the reverse diffusion process factorizes in the same way.
Details of the atom type diffusion, coordinate diffusion, and lattice diffusion are described in \cref{app:atom-type-diffusion}, \cref{app:coordinate-diffusion}, \cref{app:lattice-diffusion}, respectively. The architecture of the score network $s_\theta(\mat\tindex{t}, t)$ is described in \cref{app:architecture}. The combined objective function is presented in \cref{app:training-loss}.

\subsection{Atom type diffusion} \label{app:atom-type-diffusion}
For the diffusion of the (discrete) atom species $\types$, we use the \gls{D3PM} approach \citep{d3pm}, which is a generalization of \glspl{DDPM} to discrete data problems. As in \gls{DDPM}, the forward diffusion process is a Markov process that gradually corrupts an input sample $\type\tindex{0}$, which is a scalar discrete random variable with $K$ categories (e.g., atomic species):
\begin{equation}
    q(\type\tindex{1:T} | \type\tindex{0}) = \prod_{\tdisc = 1}^{T} q(\type\tindex{\tdisc} | \type\tindex{\tdisc - 1}),
\end{equation}
where $\type\tindex{0} \sim q(\type\tindex{0})$ is an atomic species sampled from the data distribution and $\type\tindex{T} \sim q(\type\tindex{T})$, where $q(\type\tindex{T})$ is a prior distribution that is easy to sample from.

Denoting the one-hot representation of $\type$ as a row vector $\typevec$, we can express the transitions as:
\begin{equation}
    q(\typevec\tindex{\tdisc} | \typevec\tindex{\tdisc - 1}) = \cat(\typevec\tindex{\tdisc}; \vp = \typevec\tindex{\tdisc - 1} \mQ\tindex{\tdisc}),
\end{equation}
where $[\mQ\tindex{\tdisc}]_{\nodesel \othernodesel } = q(\type\tindex{\tdisc} = \othernodesel | \type\tindex{\tdisc - 1} = \nodesel)$ is the Markov transition matrix at time step $\tdisc$. $\cat(\typevec; \vp)$ is a categorical distribution over one-hot vectors whose probabilities are given by the row vector $\vp$. Similar to \gls{DDPM}, \gls{D3PM} assumes that the forward diffusion factorizes over all discrete variables of a data point, i.e., all atomic species are diffused independently with the same transition matrices $\mQ\tindex{\tdisc}$. Hence, we only consider individual one-hot vectors in this section.
\glspl{D3PM} are trained by optimizing a variational lower bound:
\begin{align}
    L_{\textnormal{vb}} = \mathbb{E}_{q(\typevec\tindex{0})}  \Biggl [&- \mathbb{E}_{q(\typevec\tindex{1} | \typevec \tindex{0})} \log p_\vtheta(\typevec\tindex{0} | \typevec{\tindex{1}}, 1) +  \KL\left[q(\typevec\tindex{T} | \typevec\tindex{0}) \: || \: q(\typevec\tindex{T}) \right] \nonumber \\
    & + \sum_{\tdisc = 2}^{T} \mathbb{E}_{q(\typevec\tindex{\tdisc} | \typevec\tindex{0})} \KL\left[ q(\typevec\tindex{\tdisc - 1} | \typevec\tindex{\tdisc}, \typevec\tindex{0}) \: || \: p_\vtheta(\typevec\tindex{\tdisc - 1} | \typevec\tindex{\tdisc}) \right]  \Biggr ]. \label{eq:D3PM_vb}
\end{align}
Moreover, \citet{d3pm} propose an additional cross-entropy loss on the model's prediction $p_\vtheta(\typevec\tindex{0} | \typevec{\tindex{\tdisc}}, \tdisc)$:
\begin{equation*}
    L_\textnormal{CE} =  -\mathbb{E}_{q(\typevec\tindex{0})} \left[ \sum_{\tdisc = 2}^{T} \mathbb{E}_{q(\typevec\tindex{\tdisc} | \typevec\tindex{0})} \log p_\vtheta(\typevec\tindex{0} | \typevec{\tindex{\tdisc}}, \tdisc)  \right],
\end{equation*}
so that the overall loss becomes
\begin{equation}
    L = L_{\textnormal{vb}} + \lambda_\textnormal{CE} L_\textnormal{CE}.  \label{eq:d3pm_loss}
\end{equation}

Three important characteristics of \gls{DDPM} and \gls{DSM} are that (1) given $\coord\tindex{0}$ we can sample noisy samples $\coord\tindex{\tdisc}$ for arbitrary $\tdisc$ in constant time; (2) after sufficiently many diffusion steps, $\coord\tindex{T}$ follows a prior distribution that is easy to sample from; and (3) the posterior $q(\coord\tindex{\tdisc - 1} | \coord\tindex{\tdisc}, \coord\tindex{0})$ in \cref{eq:D3PM_vb} is tractable and can be computed efficiently. \gls{D3PM} also has these properties, as we briefly outline in the following:

\begin{itemize}
    \item[(1)] Fast sampling of $\typevec\tindex{\tdisc} \sim q(\typevec\tindex{\tdisc} | \typevec\tindex{0})$. Since the forward diffusion in \gls{D3PM} is governed by discrete transition matrices $\{\mQ\tindex{\tdisc}\}_{\tdisc = 1}^T$, we can write
    \begin{equation}
        q(\typevec\tindex{\tdisc} | \typevec\tindex{0}) = \cat(\typevec\tindex{\tdisc}; \vp = \typevec\tindex{\tdisc - 1} \bar{\mQ}\tindex{\tdisc}), \qquad \textnormal{where } \bar{\mQ}\tindex{\tdisc} = \mQ\tindex{1} \mQ\tindex{2}\ldots \mQ\tindex{\tdisc}.
    \end{equation}
    The cumulative transition matrices $\bar{\mQ}\tindex{\tdisc}$ can be pre-computed and for many diffusion processes even have a closed form.
    \item[(2)] Tractable prior distribution. Two of the proposed diffusion processes are the absorbing (which we employ in \mbox{MatterGen}) and uniform diffusion processes. Both gradually diffuse the data towards a limit distribution, which are the one-hot distribution on the absorbing state and the uniform distribution over all categories, respectively. For more details, we refer to Appendix A of \citet{d3pm}.
    \item[(3)] Tractable posterior  $q(\typevec\tindex{\tdisc - 1} | \typevec\tindex{\tdisc}, \typevec\tindex{0})$. Using Bayes' rule and exploiting the Markov property $q(\typevec\tindex{\tdisc} | \typevec\tindex{\tdisc - 1}, \typevec\tindex{0}) = q(\typevec\tindex{\tdisc} | \typevec\tindex{\tdisc - 1})$, we can write \begin{equation}
    q(\typevec\tindex{\tdisc - 1} | \typevec\tindex{\tdisc}, \typevec\tindex{0}) = \frac{q(\typevec\tindex{\tdisc} | \typevec\tindex{\tdisc - 1}) q(\typevec\tindex{\tdisc - 1} | \typevec\tindex{0})}{q(\typevec\tindex{\tdisc} | \typevec\tindex{0})}. \label{eq:d3pm_posterior}
    \end{equation}
    All terms in \cref{eq:d3pm_posterior} can be computed efficiently in closed form given the forward diffusion process. 
\end{itemize}

\paragraph*{Reverse sampling process.} We generate a sample $\typevec\tindex{0}$ by first sampling $\typevec\tindex{T}$ and then gradually updating it to obtain $p_\vtheta(\typevec\tindex{0:T}) = q(\typevec\tindex{T}) \prod_{\tdisc = 1}^{T} p_\vtheta(\typevec\tindex{\tdisc - 1} | \typevec\tindex{\tdisc})$. \citet{d3pm} propose to parameterize $p_\vtheta(\typevec\tindex{\tdisc - 1} | \typevec\tindex{\tdisc})$ by predicting a distribution over $\typevec\tindex{0}$ and then marginalizing it out:
\begin{equation} \label{eq:d3pm_reverse}
    p_\vtheta(\typevec\tindex{\tdisc - 1} | \typevec\tindex{\tdisc}) \propto \sum_{\typevec\tindex{0}} q(\typevec\tindex{\tdisc - 1} , \typevec{\tindex{\tdisc}} | \typevec\tindex{0}) p_\vtheta(\typevec\tindex{0} | \typevec\tindex{\tdisc}, \tdisc),
\end{equation}
where we can use our tractable posterior computation again. Since we have a discrete state space, marginalizing out $\typevec\tindex{0}$ by explicit summation has complexity $\mathcal{O}(K)$. In the case of atomic species we have $K\simeq 100$; thus, this is relatively cheap. This parameterization has the advantage that potential sparsity in the diffusion process is efficiently enforced by using $q(\typevec\tindex{\tdisc - 1}, \typevec{\tindex{\tdisc}} | \typevec\tindex{0})$ without having to be learned by the model.

\paragraph*{Forward diffusion process.} As the specific flavor of \gls{D3PM} forward diffusion we employ the masked diffusion process, which has shown best performance in the original study \citep{d3pm} as well as our initial experiments. Following \citet{d3pm}, we introduce an extra atom species [MASK] at index $K - 1$, which is the absorbing or masked state. At each timestep $\tdisc$, the transition matrices have the particularly simple form
\begin{equation}
    [\mQ\tindex{\tdisc}^\textnormal{absorbing}]_{\nodesel \othernodesel } = \begin{cases}
        1 \qquad & \textnormal{if } \nodesel = \othernodesel = m, \\
        1 - \beta\tindex{\tdisc} \qquad & \textnormal{if } \nodesel = \othernodesel \neq m, \\
        \beta\tindex{\tdisc} \qquad & \textnormal{if } \othernodesel = m \neq \nodesel, \\
        0 \qquad & \textnormal{if } m \neq \nodesel \neq \othernodesel  \neq m, 
    \end{cases}
\end{equation}
where $m$ corresponds to the absorbing state. Intuitively, each species has probability $1 - \beta\tindex{\tdisc}$ of staying unchanged, and probability $\beta\tindex{\tdisc}$ of transitioning to the absorbing state. Once a species is absorbed, it can never leave that state, and there are no transitions between different non-masked atomic species. Thus, the limit distribution of this diffusion process is a point mass on the absorbing state.

\subsection{Coordinate diffusion} \label{app:coordinate-diffusion}
For our model we perform diffusion on the \emph{fractional} coordinates and outline the approach in the following. See \cref{sec:cartesian_diffusion} for a brief outline why we favor fractional coordinate diffusion over Cartesian.
The fractional coordinates in a crystal structure live in a Riemannian manifold referred to as the flat torus $\mathbb{T}^3 = \mathbb{S}^1 \times \mathbb{S}^1 \times \mathbb{S}^1$, i.e., it is the quotient space $\mathbb{R}^3/\mathbb{Z}^3$ with equivalence relation:
\begin{equation}
    \coord + \vk \sim \coord, \qquad \vk \in \mathbb{Z}^3.
\end{equation}
Thus, adding Gaussian noise to fractional coordinates naturally corresponds to sampling from a \emph{wrapped} normal distribution, whose probability density is 
\begin{equation}
\mathcal{N}_{\textnormal{W}}\left(\bar{\coord}; \coord, \sigma^2 \mI, \mI \right) = \sum_{\vk \in \mathbb{Z}^3} \mathcal{N}\left(\bar{\coord}; \coord - \vk, \sigma^2 \mI \right), \label{eq:wrapped_normal}
\end{equation}
where $\mathcal{N}_{\textnormal{W}}\left(\vmu, \mSigma, \mB \right)$ denotes a wrapped normal distribution with mean $\vmu$, covariance matrix $\mSigma$, and periodic boundaries $\mB$. If the periodic boundaries are $[0, 1)^3$, i.e., $\mB = \mI$, we write $\mathcal{N}_{\textnormal{W}}\left(\vmu, \mSigma \right)$ for brevity.

For the diffusion of the atom coordinates we use variance-exploding diffusion, i.e., the variance of the diffusion process increases exponentially with diffusion time. This has the advantage that the prior distribution $q(\coord\tindex{T})$ is particularly simple, i.e., the uniform distribution in the range $[0, 1)^3$. \citet{torsional_diffusion} use this approach for torsional angles---which live in a 1D flat torus---in small molecule generation. The one-shot noising process of the fractional coordinates is therefore defined as 
\begin{equation}
    q(\coord\tindex{\tdisc} | \coord\tindex{0}) = \mathcal{N}_{\textnormal{W}} \left(\coord\tindex{\tdisc}; \coord\tindex{0}, \sigma\tindex{\tdisc}^2\mI \right).\label{eq:frac_coords_oneshot}
\end{equation}

\subsubsection{Variance adjustment for atomic density}
\label{subsec:variance_adjustment}

One limitation of using a constant variance for the fractional coordinate diffusion is that the diffusion in Cartesian space will have difference variance depending on the size of the unit cell. This limitation becomes clear if we express the distribution of the Cartesian coordinates $\cartcoord\tindex{\tdisc}$ using \cref{eq:frac_to_cart} via linear transformation of a Gaussian random variable $\coord\tindex{\tdisc}$:
\begin{equation}
     q(\cartcoord\tindex{\tdisc}, | \coord\tindex{0}, \lattice\tindex{\tdisc}) = \mathcal{N}_{\textnormal{W}}\left( \cartcoord\tindex{\tdisc}; \lattice\tindex{\tdisc} \coord\tindex{0}, \sigma^2\tindex{t} \lattice\tindex{\tdisc} \lattice\tindex{\tdisc}\transpose, \lattice\tindex{\tdisc} \right).  \label{eq:frac_one_shot}
\end{equation}
Observe that the covariance matrix $\mSigma\tindex{\tdisc} = \sigma^2\tindex{\tdisc} \lattice\tindex{\tdisc} \lattice\tindex{\tdisc}\transpose$ of the noisy Cartesian coordinates depends on the lattice. Thus, the (generalized) variance of the noise distribution also depends on the size of the unit cell, i.e., $|\det(\mSigma\tindex{\tdisc})|= \left(\sigma\tindex{\tdisc}^3 |\det \lattice\tindex{\tdisc}| \right)^2$. 

We mitigate this effect by scaling the variance in fractional coordinate diffusion based on the size of the unit cell. Assuming roughly constant atomic density $\mathrm{d}(\lattice\tindex{\tdisc}) = \frac{\numnodes}{\mathrm{Vol}(\lattice\tindex{\tdisc})} \propto 1 \Leftrightarrow \mathrm{Vol}(\lattice\tindex{\tdisc}) = |\det \lattice\tindex{\tdisc} | \propto \numnodes$. We therefore choose to scale $\sigma\tindex{\tdisc}$ accordingly, i.e.,
\begin{equation}
    \sigma\tindex{\tdisc}(n)=\frac{\sigma\tindex{\tdisc}}{\sqrt[3]{\numnodes}},
    \label{eq:pos_var_adjustment}
\end{equation}
such that $|\det(\mSigma\tindex{\tdisc})|= \left( \frac{\sigma^3\tindex{\tdisc}}{\numnodes} |\det \lattice\tindex{\tdisc}|\right)^2$ is no longer proportional to $\numnodes$.

\subsubsection{Score computation for fractional coordinates}
Recall from \cref{eq:score_matching_objective} that training diffusion models requires computing the score function for the one-shot transition kernel. However, for the wrapped normal distribution in \cref{eq:wrapped_normal}, (log-)likelihood and score computation are intractable because of the infinite sum. Given the thin tails of the normal distribution, both can be approximated reasonably well with a truncated sum. More specifically, the score function of the isotropic wrapped normal distribution can be expressed as
\begin{equation}
    \nabla_{\bar{\coord}} \log q_\sigma(\bar{\coord} | \coord) = \sum_{\vk \in \mathbb{Z}^3} w_\vk \frac{\bar{\coord} - \coord + \vk}{\sigma^2}, \label{eq:wn_score}
\end{equation}
where 
\begin{align}
    w_\vk =  \frac{1}{Z} \exp\left(- \frac{\|\bar{\coord} - \coord + \vk \|^2}{2\sigma^2} \right), \qquad
    Z = \sum_{\vk' \in \mathbb{Z}^3} \exp\left(- \frac{\|\bar{\coord} - \coord + \vk' \|^2}{2\sigma^2} \right).
\end{align}

\subsubsection{Fractional vs Cartesian coordinate diffusion}\label{sec:cartesian_diffusion}
Instead of diffusing fractional coordinates as in \mbox{MatterGen}, one could diffuse Cartesian coordinates, e.g., as done in CDVAE \citep{cdvae} (and in generative methods for molecules \cite{hoogeboom2022equivariant}).
 
However, this approach is not suitable for our framework.
To see this, note that while in CDVAE the lattice $\lattice$ is fixed during the diffusion of the atom coordinates, we diffuse the lattice simultaneously to the atom coordinates (and atomic species). This makes diffusion of Cartesian coordinates dependent on the lattice diffusion because the wrapped normal's covariance matrix and periodic boundaries at diffusion timestep $\tdisc$ depend on knowing the lattice matrix $\lattice\tindex{\tdisc}$. Consequently, our diffusion process from \cref{eq:diffusion_factorization} no longer factorizes into lattice and coordinates and needs to be adapted:
\begin{align}
    &q(\cartcoords\tindex{\tdisc+1}, \lattice\tindex{\tdisc+1}, \types\tindex{\tdisc+1} | \cartcoords\tindex{\tdisc}, \lattice\tindex{\tdisc}, \types\tindex{\tdisc}) \nonumber \\
    =~&q(\cartcoords\tindex{\tdisc+1} | \cartcoords\tindex{\tdisc}, \lattice\tindex{\tdisc+1}, \lattice\tindex{\tdisc})  q(\lattice\tindex{\tdisc+1} | \lattice\tindex{\tdisc}) q(\types\tindex{\tdisc+1} | \types\tindex{\tdisc}). \label{eq:diffusion_factorization_cart}
\end{align}

Here, we need to condition $q(\cartcoords\tindex{\tdisc + 1})$ on $\lattice\tindex{\tdisc + 1}$ and $\lattice\tindex{\tdisc}$ because in order to convert the Cartesian coordinates at time step $\tdisc$ to time step $\tdisc + 1$ we first need to convert $\cartcoord\tindex{\tdisc}$ to fractional coordinates using $\lattice\tindex{\tdisc}^{-1}$, and then to Cartesian coordinates at $\tdisc+1$ using $\lattice\tindex{\tdisc + 1}$. 

The one-shot distribution of noisy Cartesian coordinates (similar to \cref{eq:frac_coords_oneshot} for the fractional case) becomes:
\begin{align}    q(\cartcoord\tindex{\tdisc} | \cartcoord\tindex{0}, \{\lattice\tindex{\tdisc'}\}_{\tdisc'=1}^{\tdisc}) & = \mathcal{N}_\textnormal{W}\left(\cartcoord\tindex{\tdisc}; \lattice\tindex{\tdisc} \lattice_{0}^{-1}\cartcoord\tindex{0},\lattice\tindex{\tdisc}\left( \sum_{\tdisc'=1}^{\tdisc}\sigma\tindex{\tdisc'}^2\lattice\tindex{\tdisc'}^{-1} (\lattice\tindex{\tdisc'}\transpose)^{-1}  \right) \lattice\tindex{\tdisc}\transpose , \lattice\tindex{\tdisc} \right). 
\end{align}
Observe that we require the entire trajectory of noisy lattices $\lattice\tindex{1}, \ldots, \lattice\tindex{\tdisc}$ in order to express the noise distribution of the Cartesian atomic coordinates. This means that we first need to sample the \emph{entire} diffusion trajectory of the lattice, which is slow. Further, we have found computing the one-shot covariance matrix for the Cartesian coordinates to be numerically unstable for long diffusion trajectories. We therefore use the diffusion process of fractional coordinates described in the previous section.

\subsection{Lattice diffusion} \label{app:lattice-diffusion}
In addition to the diffusion of the atom types and coordinates described above, we also diffuse and denoise the lattice $\lattice$ in our approach. We use variance-preserving diffusion, as variance-exploding diffusion would lead to extremely large unit cells in the noisy limit. Those are challenging to handle for a \gls{GNN} with a fixed edge cutoff and would require the model to learn scores over a wide range of different length scales.

\subsubsection{Fixed-rotation lattice diffusion} \label{app:fixed_rotation_lattice_diffusion}
As the distribution of materials is invariant to global rotation, we can either choose a rotation-invariant prior distribution over unit cells, or decide on a canonical rotational alignment that we use throughout diffusion and denoising. We opt for the latter, as it gives us some more flexibility designing the diffusion process. Here, we choose to represent the lattice as a symmetric matrix. We can do so via the polar decomposition based on the singular value decomposition:
\begin{equation}
    \lattice = \mU \tilde{\lattice}, \qquad \mU = \mW \mV\transpose, \qquad \tilde{\lattice} = \mV \mSigma \mV\transpose,
\end{equation}
where $\mW$ and $\mV$ are the left and right singular vectors of $\lattice$, respectively, and $\mSigma$ is the diagonal matrix of singular values. $\mU$ is a rotation matrix and $\tilde{\lattice}$ is a symmetric positive-definite matrix. 

We restrict our entire forward lattice diffusion to symmetric matrices by enforcing the noise $\vz \in \mathbb{R}^{3\times 3}$ on the lattice to be symmetric, e.g., by only modeling the upper-triangular part of the matrix and mirroring it to the lower triangular part. Notice that this effectively fixes the rotation, i.e., we only have six degrees of freedom left. Going forward, we only consider symmetric lattices and lattice noise.

\subsubsection{Lattice diffusion with custom limit mean and variance}
Naively using variance-preserving diffusion independently on the lattice vectors leads to the following forward diffusion distribution:
\begin{equation}
    q(\lattice\tindex{\tdisc} | \lattice\tindex{0}) = \mathcal{N}\left(\sqrt{\bar{\alpha}\tindex{\tdisc}}\lattice\tindex{0}, (1 - \bar{\alpha}\tindex{\tdisc}) \mI \right).
\end{equation}
However, for large $\tdisc$ we observed that the majority of the resulting unit cells have very small volume and steep angles, which means that the atoms are extremely densely packed inside the noisy cells. We therefore modify the diffusion process as follows:
\begin{equation}
    q(\lattice\tindex{\tdisc} | \lattice\tindex{0}) = \mathcal{N}\left(\sqrt{\bar{\alpha}\tindex{\tdisc}}\lattice\tindex{0} + (1 - \sqrt{\bar{\alpha}\tindex{\tdisc}}) \mu(\numnodes)\mI, (1-\bar{\alpha}\tindex{\tdisc})\sigma\tindex{\tdisc}^2(\numnodes) \mI \right), \label{eq:lattice_diffusion}
\end{equation}
which yields the following limit distribution for $T \rightarrow \infty$:
\begin{equation}
    q(\lattice\tindex{T}) = \mathcal{N}\left( \mu(\numnodes)\mI, \sigma\tindex{T}^2(\numnodes) \mI \right).
\end{equation}
Thus, in the limit distribution we have a tendency towards cubic lattices $\lattice \propto \mI$, which often occur in nature and have a relatively narrow range of volumes. Further, the lattice vector angles when sampling from the prior are mostly concentrated between 60° and 120°. This aligns both with the range of angles of Niggli-reduced cells as well as the initialization range of cell vector angles in \gls{RSS} \citep{airss}, suggesting that this is a good starting point for the generation process. 

To understand the choice of the mean in the limit distribution, recall that the volume of a parallelepiped $\lattice$ is $|\det \lattice|$. 
We introduce a scalar coefficient $\mu(\numnodes)$ that depends on the number of atoms in the cell to make the atomic density of the mean noisy lattice roughly constant for differently sized systems. Setting $\mu(\numnodes) = \sqrt[3]{\numnodes c}$, the volume of the prior mean becomes $\mathrm{Vol}(\sqrt[3]{\numnodes  c} \mI) = \numnodes c$. Thus, the atomic density of the prior mean becomes $\mathrm{d}(\mu(\numnodes)\mI) = \frac{\numnodes}{\mathrm{Vol}(\mu(\numnodes)\mI)}=\frac{1}{c}$. We can set $c$ to the inverse average density of the dataset as a reasonable prior.

Similarly, we adjust the variance in the limit distribution to be proportional to the volume, such that the signal-to-noise ratio of the noisy lattices is constant across the numbers of nodes. To this end, we set the limit standard deviation as $\sigma(\numnodes) = \sqrt[3]{\numnodes\nu}$. Thus, for a diagonal entry of the lattice matrix the signal-to-noise-ratio in the limit is $\lim_{t\rightarrow \infty}\frac{|\mu(\numnodes)|}{\sigma(\numnodes)} = \frac{\sqrt[3]{\numnodes c}}{\sqrt[3]{\numnodes\nu}} = \left(\frac{c}{\nu}\right)^{1/3}$ and therefore independent of the number of atoms.

\subsection{Architecture of the score network} \label{app:architecture}
We employ an SE(3)-equivariant \gls{GNN} to predict scores for the lattice, atom positions, and atom types in the denoising process.
In particular, we adapt the GemNet architecture by \citet{gasteiger2021gemnet}, originally developed to be a universal \gls{MLFF}.
GemNet is a symmetric message-passing \gls{GNN} that uses directional information to achieve SO(3)-equivariance, and incorporates two- and three-body information in the first layer for efficiency.
Since we do not predict energies, we adapt the direct (i.e., non-conservative) force prediction variant of the model, named GemNet-dT, which has been shown to be more computationally efficient and accurate in these scenarios \cite{gasteiger2021gemnet}.
We employ four message-passing layers, a cutoff radius of 7~$\text{\AA}$ for the neighbor list construction, and set the dimension of hidden representations for nodes and edges to 512.

We train the model to predict Cartesian coordinate scores $\vs_{\coords, \vtheta}(\coords\tindex{\tdisc}, \lattice\tindex{\tdisc}, \types\tindex{\tdisc}, \tdisc)$ as if they were non-conservative forces.
These scores are equivariant to rotation and permutation, and invariant to translation.
We then transform the Cartesian scores into fractional scores following \cref{eq:frac_to_cart}.

For the atom-type reverse diffusion, recall from \cref{eq:d3pm_reverse} that we predict the atom types $\types\tindex{0}$ given the atom types $\types\tindex{\tdisc}$ at time $\tdisc$. For materials, we additionally condition on lattice $\lattice\tindex{\tdisc}$ and coordinates $\coords\tindex{\tdisc}$; more precisely, the (unnormalized) log-probabilities $\log p_\vtheta(\types\tindex{0} | \coords\tindex{\tdisc}, \lattice\tindex{\tdisc}, \types\tindex{\tdisc}, \tdisc)$ of the atomic species at $\tdisc = 0$ 
are computed as:
\begin{equation}
\log p_\vtheta(\types\tindex{0} | \coords\tindex{\tdisc}, \lattice\tindex{\tdisc}, \types\tindex{\tdisc}, \tdisc) =  \mH^{(L)} \mW,
\label{eq:atom_type_score_predictions}
\end{equation}
where $\mH^{(L)} \in \mathbb{R}^{\numnodes \times \hiddendim}$ are the hidden representations of atoms at the last message-passing layer $L$, and $\mW \in \mathbb{R}^{\hiddendim \times K}$ are the weights of a fully-connected linear layer, with $K$ being the number of atom types (including the masked null state).

\subsubsection{Computation of the predicted lattice scores}
\label{sec:lattice_scores}
To predict the lattice scores $\vs_{\lattice, \vtheta}(\coords\tindex{\tdisc}, \lattice\tindex{\tdisc}, \types\tindex{\tdisc}, \tdisc)$, we utilize the model's hidden representations of the edges. For layer $l$, we denote the edge representation of the edge $(\nodesel \othernodesel \vk) \in \edgeset$ between atoms $\nodesel$ and $\othernodesel$ as $\vm_{\nodesel \othernodesel \vk}^l \in \mathbb{R}^{\hiddendim}$, where $\nodesel$ is inside the unit cell and $\othernodesel$ is $\vk \in \mathbb{Z}^3$ unit cells displaced from the center unit cell. We use a \gls{MLP} $\phi^l: \mathbb{R}^\hiddendim \rightarrow \mathbb{R}$ to predict a scalar score per edge. We treat this as a prediction by the model indicating by how much an edge's length should increase or decrease, and translate this into a predicted transformation of the lattice via chain rule derivation:
\begin{align}
    \frac{\partial \tilde{d}_{\nodesel \othernodesel\vk}}{\partial \lattice\tindex{\tdisc}} &= \frac{\partial}{\partial \lattice\tindex{\tdisc}} \left \|\lattice\tindex{\tdisc} \left(\coord\tindex{\tdisc}\nindex{\othernodesel} - \coord\tindex{\tdisc}\nindex{\nodesel} + \vk \right) \right \|_2 \nonumber \\
    &= \frac{1}{\tilde{d}_{\nodesel \othernodesel\vk}} \lattice\tindex{\tdisc} \left(\coord\tindex{\tdisc}\nindex{\othernodesel} - \coord\tindex{\tdisc}\nindex{\nodesel} + \vk \right) \cdot \left(\coord\tindex{\tdisc}\nindex{\othernodesel} - \coord\tindex{\tdisc}\nindex{\nodesel} + \vk \right)\transpose \nonumber \\
    &= \frac{1}{\tilde{d}_{\nodesel \othernodesel \vk}} \tilde{\vd}_{\nodesel \othernodesel \vk}(\vd_{\nodesel \othernodesel\vk})\transpose,
\end{align}
where $\tilde{d}_{\nodesel \othernodesel \vk}= \|\tilde{\vd}_{\nodesel \othernodesel\vk}\|_2$ and $\tilde{\vd}_{\nodesel \othernodesel \vk} = \lattice\tindex{\tdisc} \left(\coord\tindex{\tdisc}\nindex{\othernodesel} - \coord\tindex{\tdisc}\nindex{\nodesel} + \vk \right)$ are the edge length and edge displacement in Cartesian coordinates, respectively, and $\vd_{\nodesel \othernodesel\vk} = \coord\tindex{\tdisc}\nindex{\othernodesel} - \coord\tindex{\tdisc}\nindex{\nodesel} + \vk$ is the edge displacement in fractional coordinates. The predicted lattice score \emph{per edge} is then $\phi^l(\vm_{\nodesel \othernodesel \vk}^l) \cdot \frac{\partial \tilde{d}_{\nodesel \othernodesel \vk}}{\partial \lattice\tindex{\tdisc}}$. These predicted scores are averaged over all edges $(\nodesel \othernodesel \vk) \in \edgeset$ to get the predicted lattice score for layer $l$:
\begin{equation}
    \hat{\vs}_{\lattice, \vtheta}^l(\coords\tindex{\tdisc}, \lattice\tindex{\tdisc}, \types\tindex{\tdisc}, \tdisc) = \frac{1}{|\edgeset|} \sum_{(\nodesel \othernodesel \vk) \in \edgeset} \phi^l(\vm_{\nodesel \othernodesel \vk}^l) \cdot \frac{1}{\tilde{d}_{\nodesel \othernodesel \vk}} \tilde{\vd}_{\nodesel \othernodesel \vk}(\vd_{\nodesel \othernodesel \vk})\transpose.
\end{equation}

Stacking the model's predictions into a diagonal matrix $\mPhi^l \in \mathbb{R}^{|\edgeset| \times |\edgeset|} = \mathrm{diag}\left(\frac{\phi^l(\vm_{\nodesel \othernodesel \vk}^l)}{|\edgeset| \cdot \tilde{d}_{\nodesel \othernodesel \vk}} \right)$, we can write more concisely
\begin{equation}
    \hat{\vs}_{\lattice, \vtheta}^l(\coords\tindex{\tdisc}, \lattice\tindex{\tdisc}, \types\tindex{\tdisc}, \tdisc) = \tilde{\mD} \mPhi^l \mD\transpose = \lattice\tindex{\tdisc} \mD \mPhi^l \mD\transpose,
\end{equation}
where $\tilde{\mD}, \mD \in \mathbb{R}^{3 \times |\edgeset|}$ are the stacked matrices of Cartesian and fractional distance vectors, respectively, with $\tilde{\mD} = \lattice_i \mD$ for structure $i$.
This form reveals that these predicted lattice scores have a key shortcoming. To see this, recall from \cref{app:fixed_rotation_lattice_diffusion} that we perform lattice diffusion on the subspace of \emph{symmetric} lattice matrices.
However, the lattice scores from $\hat{\vs}_{\lattice, \vtheta}^l(\coords\tindex{\tdisc}, \lattice\tindex{\tdisc}, \types\tindex{\tdisc}, \tdisc)$ are generally not symmetric matrices.
We address this issue with the following modification:
\begin{align}
    \vs_{\lattice, \vtheta}^l(\coords\tindex{\tdisc}, \lattice\tindex{\tdisc}, \types\tindex{\tdisc}, \tdisc) &=  \tilde{\vs}_{\lattice, \vtheta}^l(\coords\tindex{\tdisc}, \lattice\tindex{\tdisc}, \types\tindex{\tdisc}, \tdisc)\lattice\tindex{\tdisc}\transpose \nonumber \\
    &= \lattice\tindex{\tdisc} \mD \tilde{\mPhi}^l \mD\transpose \lattice\tindex{\tdisc}\transpose = \tilde{\mD} \tilde{\mPhi}^l \tilde{\mD}\transpose,
    \label{eq:symmetric_lattice_score}
\end{align}
where $\tilde{\mPhi}^l = \mathrm{diag}\left(\frac{\phi^l(\vm_{\nodesel \othernodesel \vk}^l)}{|\edgeset| \cdot d^2_{\nodesel \othernodesel \vk}} \right)$.
Finally, the predicted layer-wise lattice scores are summed to obtain the predicted lattice score: 
\begin{equation} \label{eq:lattice_score_predictions}
    \vs_{\lattice, \vtheta}(\coords\tindex{\tdisc}, \lattice\tindex{\tdisc}, \types\tindex{\tdisc}, \tdisc) = \sum_{l=1}^L \vs_{\lattice, \vtheta}^l(\coords\tindex{\tdisc}, \lattice\tindex{\tdisc}, \types\tindex{\tdisc}, \tdisc),
\end{equation}
which is scale-invariant and equivariant under rotation. 
The equivariance derives from the way it is composed with the Cartesian coordinate matrix, and the scale invariance is due to the normalization happening inside $\tilde{\mPhi}^l$.
In particular, the diagonal entries of $\tilde{\mPhi}^l$ related to the edges are normalized three times: they are divided by the total number of edges, and then multiplied twice by the inverse of the norm of the edge vectors.
Given these properties, $ \hat{\vs}_{\vtheta}$ behaves like a symmetric stress tensor $\vsigma$, since the stress tensor is scale-invariant and equivariant under the rotation operator $\mR$:
\begin{align}
\vsigma ' (\lambda \mat) &= \vsigma(\mat), \\
\vsigma ' (\mR \mat) &= \mR \vsigma (\mR \mat) \mR\transpose,
\end{align}
where we use $\lambda$ to indicate the supercell replication operation.

\subsubsection{Augmenting the input with lattice information}
\label{sec:inserting_lattice_info}
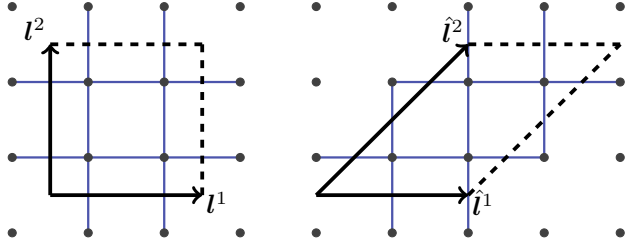
\begin{figure}[th]
    \centering
\begin{tikzpicture} 
\definecolor{edgecolor}{RGB}{76,85,172}
\node at (0.5,0.5) (O) {};
\node at (0.3,2.7) (l2) {$\lvec^2$};
\node at (2.7,0.4) (l1) {$\lvec^1$};
\node at (5.8,2.7) (l2) {$\hat{\lvec}^2$};
\node at (6.2,0.4) (l1) {$\hat{\lvec}^1$};

\draw [edgecolor,line width=0.3mm] (1.0, 0.0) -- (1.0, 1.0);
\draw [edgecolor,line width=0.3mm] (1.0, 1.0) -- (1.0, 2.0);
\draw [edgecolor,line width=0.3mm] (1.0, 2.0) -- (1.0, 3.0);
\draw [edgecolor,line width=0.3mm] (1.0, 1.0) -- (0.0, 1.0);
\draw [edgecolor,line width=0.3mm] (1.0, 1.0) -- (2.0, 1.0);
\draw [edgecolor,line width=0.3mm] (2.0, 0.0) -- (2.0, 1.0);
\draw [edgecolor,line width=0.3mm] (3.0, 1.0) -- (2.0, 1.0);
\draw [edgecolor,line width=0.3mm] (2.0, 2.0) -- (3.0, 2.0);
\draw [edgecolor,line width=0.3mm] (2.0, 2.0) -- (2.0, 1.0);
\draw [edgecolor,line width=0.3mm] (2.0, 2.0) -- (2.0, 3.0);
\draw [edgecolor,line width=0.3mm] (2.0, 2.0) -- (1.0, 2.0);
\draw [edgecolor,line width=0.3mm] (1.0, 2.0) -- (0.0, 2.0);

\draw [edgecolor,line width=0.3mm] (5.0, 1.0) -- (5.0, 2.0);
\draw [edgecolor,line width=0.3mm] (5.0, 1.0) -- (5.0, 0.0);
\draw [edgecolor,line width=0.3mm] (5.0, 1.0) -- (6.0, 1.0);
\draw [edgecolor,line width=0.3mm] (5.0, 1.0) -- (4.0, 1.0);
\draw [edgecolor,line width=0.3mm] (6.0, 1.0) -- (7.0, 1.0);
\draw [edgecolor,line width=0.3mm] (6.0, 1.0) -- (6.0, 2.0);
\draw [edgecolor,line width=0.3mm] (6.0, 1.0) -- (6.0, 0.0);
\draw [edgecolor,line width=0.3mm] (6.0, 2.0) -- (7.0, 2.0);
\draw [edgecolor,line width=0.3mm] (6.0, 2.0) -- (5.0, 2.0);
\draw [edgecolor,line width=0.3mm] (6.0, 2.0) -- (6.0, 3.0);
\draw [edgecolor,line width=0.3mm] (7.0, 2.0) -- (7.0, 3.0);
\draw [edgecolor,line width=0.3mm] (7.0, 2.0) -- (8.0, 2.0);
\draw [edgecolor,line width=0.3mm] (7.0, 2.0) -- (7.0, 1.0);

\draw [-to, line width=0.5mm] (0.5, 0.5) -- (0.5, 2.5);
\draw [-to, line width=0.5mm] (0.5, 0.5) -- (2.5, 0.5);
\draw [dashed, line width=0.5mm] (0.5, 2.5) -- (2.5, 2.5);
\draw [dashed, line width=0.5mm] (2.5, 2.5) -- (2.5, 0.5);

\draw [-to, line width=0.5mm] (4.0, 0.5) -- (6.0, 2.5);
\draw [-to, line width=0.5mm] (4.0, 0.5) -- (6.0, 0.5);
\draw [dashed, line width=0.5mm] (6.0, 2.5) -- (8.0, 2.5);
\draw [dashed, line width=0.5mm] (6.0, 0.5) -- (8.0, 2.5);

\foreach \x in {0,...,8} 
{ 
  \foreach \y in {0,...,3} 
  { 
    \filldraw [darkgray] (\x,\y) circle [radius=1.5pt]; 
  }; 
}; 
\end{tikzpicture}
\vspace{0.3cm}
    \caption{Diagram showing two equivalent lattice choices $\lattice$ and $\hat{\lattice} = \lattice \mC$ that lead to the same periodic structure. The dots represent atoms in the 2D periodic structures, and the blue lines indicate edges of atoms inside the unit cell to their four nearest neighbors, inside and outside the unit cell. Note that both choices of unit cell lead to indistinguishable structures, as indicated by the identical placement of atoms and equivalent edges. Here, $\mC =  \left[\begin{array}{ccc}
        1 & 1 \\
        0 & 1 \\
    \end{array}\right]$.}
    
\label{fig:equivalent_lattices}
\end{figure}

The chain-rule-based lattice score predictions from \cref{eq:lattice_score_predictions} have shown to lack expressiveness for modeling the score of our Gaussian forward diffusion in our early experiments. We hypothesize that this is because our periodic \gls{GNN} model is invariant to the particular choice of unit cell. For instance, it cannot distinguish the two structures in \cref{fig:equivalent_lattices}. To address this, we drop the invariance of the \gls{GNN} w.r.t. equivalent choices of the unit cell by injecting information about the lattice angles into the internal representations. This means that the generative distribution is no longer invariant to the concrete choice of unit cell.  
We nonetheless note that any lattice can be \textit{uniquely} transformed into its so-called Niggli-reduced cell \citep{niggli}.
We apply this transformation to all training data points and, consequently, side-step the loss of cell choice equivariance we introduce.
Concretely, we concatenate the roto-translation invariant input edge representations $\vm^{\textnormal{inp}}$ with the cosines of the angles of the edge vectors w.r.t. the lattice cell vectors:
\begin{equation}
    \hat{\vm}_{ij\vk}^{\textnormal{inp}} = \left( \vm_{ij\vk}^{\textnormal{inp}}, \; \cos(\vd_{ij\vk}, \lvec\nindex{1}), \; \cos(\vd_{ij\vk}, \lvec\nindex{2}) , \; \cos(\vd_{ij\vk}, \lvec\nindex{3}) \right).
    \label{eq:lattice_info_addition}
\end{equation}
This additional information allows the model to distinguish the two cases in \cref{fig:equivalent_lattices}, while the internal representations remain invariant to rotation and translation.

\subsection{Training loss}
\label{app:training-loss}
Our model is trained to minimize the following loss, which is a sum of the score matching loss (see \cref{eq:score_matching_objective}) for the coordinates and cell, respectively, and the atom type loss (compare with \gls{D3PM} objective in \cref{eq:d3pm_loss}):
\begin{equation}
    L = \lambda_\textnormal{coord} L_\textnormal{coord} + \lambda_\textnormal{cell} L_\textnormal{cell} + \lambda_\textnormal{types}L_\textnormal{types}, \label{eq:training_loss}
\end{equation}
where 
\begin{equation}
    L_\textnormal{coord} = \sum_{\tdisc=1}^T \sigma\tindex{\tdisc}(\numnodes)^2 \mathbb{E}_{q(\coord\tindex{0})} \mathbb{E}_{q(\coord\tindex{\tdisc} | \coord\tindex{0})}\left[\|\vs_{\coord, \vtheta}(\coords\tindex{\tdisc}, \lattice\tindex{\tdisc}, \types\tindex{\tdisc}, \tdisc) - \nabla_{\coord\tindex{\tdisc}}\log q(\coord\tindex{\tdisc} | \coord\tindex{0}) \|_2^2 \right], \label{eq:coord_loss}
\end{equation}

\begin{equation}
    L_\textnormal{cell} = \sum_{\tdisc=1}^T (1-\bar{\alpha}\tindex{\tdisc})\sigma\tindex{\tdisc}(\numnodes)^2 \mathbb{E}_{q(\lattice\tindex{0})} \mathbb{E}_{q(\lattice\tindex{\tdisc} | \lattice\tindex{0})}\left[\|\vs_{\lattice, \vtheta}(\coords\tindex{\tdisc}, \lattice\tindex{\tdisc}, \types\tindex{\tdisc}, \tdisc) - \nabla_{\lattice\tindex{\tdisc}}\log q(\lattice\tindex{\tdisc} | \lattice\tindex{0}) \|_2^2 \right] \label{eq:cell_loss}
\end{equation}

\begin{align}
    L_{\textnormal{types}} = \mathbb{E}_{q(\typevec\tindex{0})} & \Biggl [ \sum_{\tdisc = 2}^{T} \mathbb{E}_{q(\typevec\tindex{\tdisc} | \typevec\tindex{0})}[ \KL\left[ q(\typevec\tindex{\tdisc - 1} | \typevec\tindex{\tdisc}, \typevec\tindex{0}) \: || \: p_\vtheta(\typevec\tindex{\tdisc - 1} |  \coords\tindex{\tdisc}, \lattice\tindex{\tdisc}, \types\tindex{\tdisc}) \right] \notag  \\
    & - \lambda_{\textnormal{CE}}\log p_\vtheta(\typevec\tindex{0} |  \coords\tindex{\tdisc}, \lattice\tindex{\tdisc}, \types\tindex{\tdisc}, \tdisc)]   - \mathbb{E}_{q(\typevec\tindex{1} | \typevec \tindex{0})} \left[\log p_\vtheta(\typevec\tindex{0} |  \coords\tindex{1}, \lattice\tindex{1}, \types\tindex{1}, 1) \right] \Biggr ]. \label{eq:types_loss}
\end{align}

For simplicity, \cref{eq:coord_loss,eq:types_loss} show the loss only for a single atom's coordinates and specie, respectively; the overall losses for coordinates and atom types sum over all atoms in a structure.

\newpage
\section{Fine-tuning the score network for property-guided generation}\label{app:fine-tuning}

Here we discuss our fine-tuning procedure of the score model to enable \emph{property-guided} generation via classifier-free guidance \citep{ho2022classifier}.

\subsection{Fine-tuning the score network with adapter modules}
\label{sec:fine_tuning}

Leveraging the large-scale unlabeled \gls{Alex-MP-20} dataset enables \mbox{MatterGen} to generate a broad distribution of stable material structures via reverse diffusion, driven by unconditional scores. To facilitate conditional generation with classifier-free guidance, the property-conditioned scores need to be learned through a labeled dataset. However, labeled datasets, often limited in size and diversity, present challenges in learning the conditional scores from scratch.

To enable rapid learning of the conditional scores in the sparsely-labeled data regime, we propose to fine-tune the unconditional score network with additional trainable adapter modules. Each adapter layer is a combination of an \gls{MLP} layer and a zero-initialized mix-in layer~\citep{zhang2023adding}, so \mbox{MatterGen} still outputs the learned unconditional scores at initialization. This is desired because the unconditional scores have been optimized to generate stable materials during pre-training, which is a prerequisite for modeling the property-conditional distribution of materials.

The additional adapter modules consist of an embedding layer $f_{\mathrm{embed}}$ for the property label that outputs a property embedding $\vg$, and a series of adapter layers, one before each message-passing layer (four in total). The adapter layer augments the atom embedding of the original GemNet score network to incorporate property information. Concretely, at the $L$-th interaction layer, given the property embedding $\vg$ and the intermediate node hidden representation $\{\mH^{(L)}_j\}_{j=1}^n$, the property-augmented node hidden representation $\{\mH'^{(L)}_j\}_{j=1}^n$ is given by:
\begin{equation}
    \mH'^{(L)}_j = \mH^{(L)}_j + 
 f^{(L)}_{\mathrm{mixin}} \left( f^{(L)}_{\mathrm{adapter}}\left( \vg \right) \right) \cdot \mathbb{I}(\text{property is not null}),
\end{equation}
where $f^{(L)}_{\mathrm{mixin}}$ is the $L$-th mix-in layer, which is a zero-initialized linear layer without bias weights. 
$f^{(L)}_{\mathrm{adapter}}$ is the $L$-th adapter layer, which is a two-layer \gls{MLP} model. The indicator function $\mathbb{I}(\text{property is not null})$ ensures the model outputs the unconditional score when no conditional label is given. The adapter modules add additional weights of $f_{\mathrm{embed}}$,  $f^{(L)}_{\mathrm{adapter}}$, and $f^{(L)}_{\mathrm{mixin}}$ to each layer of the model. 
 
The fine-tuning process uses the same training objective as the unconditional pre-training stage with conditional property labels incorporated. When fine-tuning finishes, the score network is able to predict both conditional and unconditional scores. The fine-tuned model enables us to generate structures satisfying the property condition without a major sacrifice in terms of stability and novelty. With the unconditional score network as a strong initialization, the fine-tuning procedure is more computation- and sample-efficient than re-training from scratch if the labeled dataset is only sparsely labeled. 

\subsection{Classifier-free guidance}
To generate samples conditioned on a specific value $c$ of a property, we adopt classifier-free diffusion guidance \cite{ho2022classifier} throughout this work. In classifier-free guidance, a guidance factor $\gamma$ is applied to the conditional distribution $p(\mat\tindex{\tdisc}|c)$, such that

\begin{equation}\label{eqn:appendix_E_classifier_free}
    \begin{split}
        p_{\gamma}(\mat\tindex{\tdisc}|c)&\propto p(c|\mat\tindex{\tdisc})^{\gamma}p(\mat\tindex{\tdisc}) \\
        &\propto \left(\frac{p(\mat\tindex{\tdisc}|c)}{p(\mat\tindex{\tdisc})}\right)^{\gamma} p(\mat\tindex{\tdisc}) \\
        &\propto p(\mat\tindex{\tdisc}|c)^{\gamma} p(\mat\tindex{\tdisc})^{1-\gamma}
    \end{split}
\end{equation}
is used instead of $p(\mat\tindex{\tdisc})$ when evaluating the model score during the reverse process in the conditional setting. We adopt a value of $\gamma=2$ in all conditional generation experiments.

\subsubsection{Continuous case}
\label{subsec:continuous_embedding}

The conditional score follows from \cref{eqn:appendix_E_classifier_free} by taking gradients of the logarithm w.r.t. continuous variables in $\mat\tindex{\tdisc}$. For example, for fractional coordinates $\coords\tindex{\tdisc}$ we have

\begin{equation}
    \nabla_{\coords\tindex{\tdisc}} \ln p_{\gamma}(\coords\tindex{\tdisc}|c) = \gamma \nabla_{\coords\tindex{\tdisc}} \ln q(\coords\tindex{\tdisc}|c) + (1-\gamma)\nabla_{\coords\tindex{\tdisc}} \ln q(\coords\tindex{\tdisc}).
\end{equation}
Practically, learning a conditional score $\nabla_{\coords\tindex{\tdisc}} \ln q(\coords\tindex{\tdisc}|c)$ equates to concatenating a latent embedding $\vg_c \in \mathbb{R}^d$ of the condition $c$ to the score model $\vs_\vtheta(\mat\tindex{\tdisc}, \vg_c, \tdisc)$ during score matching. The unconditional score $\nabla_{\coords\tindex{\tdisc}} \ln p(\coords\tindex{\tdisc})$ is obtained by providing a null embedding for the condition, i.e., using $\vs_\vtheta(\mat\tindex{\tdisc}, \vg_c=\mathrm{null}, \tdisc)$. When we condition on multiple properties, the conditional score for $N$ properties with embeddings $\vg_{c_{i}}$ is obtained by $\vs_\vtheta(\mat\tindex{\tdisc}, \vg_{c_{1}}, \vg_{c_{2}},\ldots,\vg_{c_N}, \tdisc)$.

\subsubsection{Discrete case}
\label{subsec:discrete_embedding}
The model's task in denoising discrete atom types $\typevec$ is to fit and predict $\tilde{q}(\typevec\tindex{\tdisc - 1} | \typevec\tindex{\tdisc}, c)$. Following Eq.~(4) in \citep{d3pm}, we can rewrite this as
\begin{align*}
    \tilde{q}(\typevec\tindex{\tdisc - 1} | \typevec\tindex{\tdisc}, c) &\propto  \sum_{\typevec\tindex{0}} q(\typevec\tindex{\tdisc - 1}, \typevec\tindex{\tdisc}| \typevec\tindex{0}) \cdot \tilde{q}(\typevec\tindex{0} | \typevec\tindex{\tdisc}, c).
\end{align*}
Thus, the predictive task is to approximate $p_\theta(\typevec\tindex{0} | \typevec\tindex{\tdisc}, c) \approx \tilde{q}(\typevec\tindex{0} | \typevec\tindex{\tdisc}, c)$. For this distribution we can perform classifier-free guidance as follows:
\begin{align*}
    \tilde{q}_\gamma(\typevec\tindex{0} | \typevec\tindex{\tdisc}, c) &\propto \tilde{q}(c | \typevec\tindex{0}, \typevec\tindex{\tdisc})^\gamma \cdot \tilde{q}(\typevec\tindex{0} | \typevec\tindex{\tdisc}) \\
    &= \left(\frac{\tilde{q}(\typevec\tindex{0} | c, \typevec\tindex{\tdisc}) \cdot \tilde{q}(c | \typevec\tindex{\tdisc})}{\tilde{q}(\typevec\tindex{0} | \typevec\tindex{\tdisc})} \right)^\gamma \cdot \tilde{q}(\typevec\tindex{0} | \typevec\tindex{\tdisc}) \\
    &\propto \tilde{q}(\typevec\tindex{0} | c, \typevec\tindex{\tdisc})^\gamma \cdot \tilde{q}(\typevec\tindex{0} | \typevec\tindex{\tdisc})^{1-\gamma}.
\end{align*}
We can approximate this guided distribution accordingly with an unconditional and a conditional prediction model, i.e., $ p_\theta(\typevec\tindex{0} | c, \typevec\tindex{\tdisc}, \tdisc)^\gamma \cdot p_\theta(\typevec\tindex{0} | \typevec\tindex{\tdisc}, \tdisc)^{1-\gamma} \approx \tilde{q}(\typevec\tindex{0} | c, \typevec\tindex{\tdisc})^\gamma \cdot \tilde{q}(\typevec\tindex{0} | \typevec\tindex{\tdisc})^{1-\gamma}$.
Taking the logarithm, we obtain
\begin{equation*}
    \log \left( p_\theta(\typevec\tindex{0} | c, \typevec\tindex{\tdisc}, \tdisc)^\gamma \cdot p_\theta(\typevec\tindex{0} | \typevec\tindex{\tdisc}, \tdisc)^{1-\gamma} \right) = 
    \gamma \log p_\theta(\typevec\tindex{0} | c, \typevec\tindex{\tdisc}, \tdisc) + (1-\gamma) \log p_\theta(\typevec\tindex{0} | \typevec\tindex{\tdisc}, \tdisc).
\end{equation*}

\section{Dataset generation}\label{app:data}
Here we provide details about the training dataset \gls{Alex-MP-20} and the reference dataset \gls{Alex-MP-ICSD} used throughout this work.
\subsection{Data sources}\label{sec:data}

We obtained crystal structures via three sources:

\begin{itemize}
    \item \gls{MP} (v2022.10.28, Creative Commons Attribution 4.0 International License) \cite{jain2013commentary}, an open-access resource containing \gls{DFT}-relaxed crystal structures obtained from a variety of sources, but largely based upon experimentally-known crystals.
    \item The \gls{Alex} dataset \cite{alexandria,Schmidt_2021,Schmidt_2023} (Creative Commons Attribution 4.0 International License), an open-access resource containing \gls{DFT}-relaxed crystal structures from a variety of sources, including a large quantity of hypothetical crystal structures generated by ML methods or other algorithmic means.
    \item \gls{ICSD} (release 2023.1) \cite{Zagorac2019}, a proprietary database containing crystal structures refined from experiment. For the purposes of dataset generation, we queried \gls{ICSD} only for experimental crystal structures that were not tagged as already included in \gls{MP}, and that were directly calculable by \gls{DFT} (i.e., ordered crystals).
\end{itemize}

\begin{figure}[th]
    \centering
    \includegraphics[width=\textwidth]{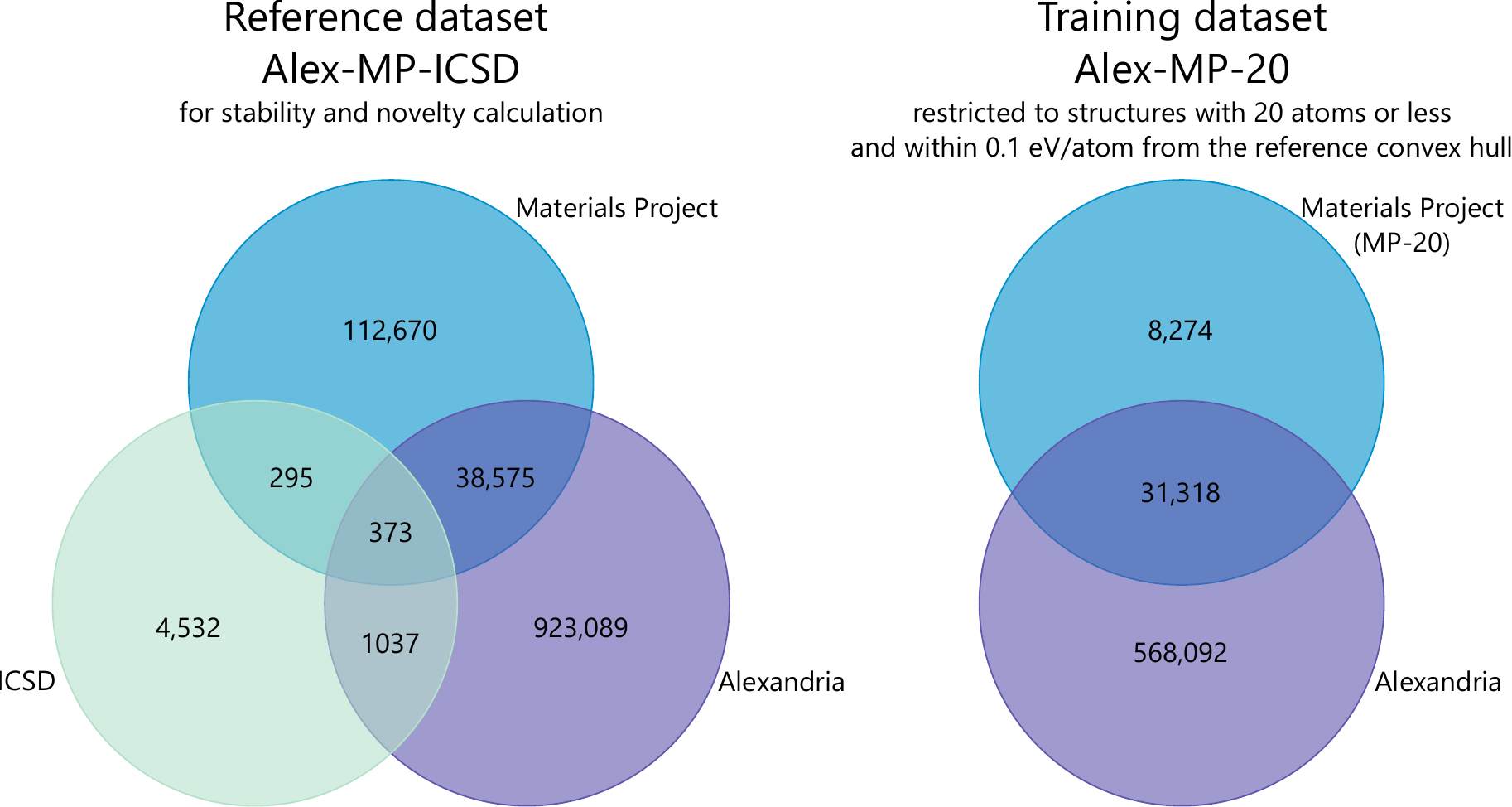}
    \caption{Venn diagrams (not to scale) showing the overlap of crystal structures from the three data sources used in this work, for the reference \gls{Alex-MP-ICSD} dataset (left) and the \gls{Alex-MP-20} training dataset (right).
    Crystal structures were de-duplicated after calculation and therefore the overlap in this diagram shows cases where the same crystal structure was present in multiple data sources. Note that the statistics for \gls{ICSD} include only the structures sourced from the \gls{ICSD} in this study, and not the full \gls{ICSD} database.}
    \label{fig:datasets-venn}
\end{figure}

For crystal structures from \gls{MP}, we retrieved existing calculations via the \gls{MP} API. For other data sources, we performed new calculations using \gls{MP} settings to guarantee consistency of data (see \cref{sec:dft_details}). We then followed \gls{MP}'s data analysis approach as implemented in \texttt{emmet} \cite{emmet}, which includes the following steps:

\begin{enumerate}
    \item Validation of each individual \gls{DFT} calculation to ensure required minimum quality criteria are met.
    \item Grouping of calculations of equivalent crystal structures, which de-duplicates crystal structures when the same crystal is present in multiple data sources. See \cref{fig:datasets-venn} for an overview of the resulting statistics.
    \item Application of an empirical correction scheme \cite{wang2021framework} to address known systematic errors from the \gls{PBE} functional.
    \item Construction of convex hull phase diagrams for each chemical system.
\end{enumerate}

\begin{figure}[ht!]
     \centering
     \begin{subfigure}[b]{0.8\textwidth}
         \centering
         \includegraphics[width=\textwidth]{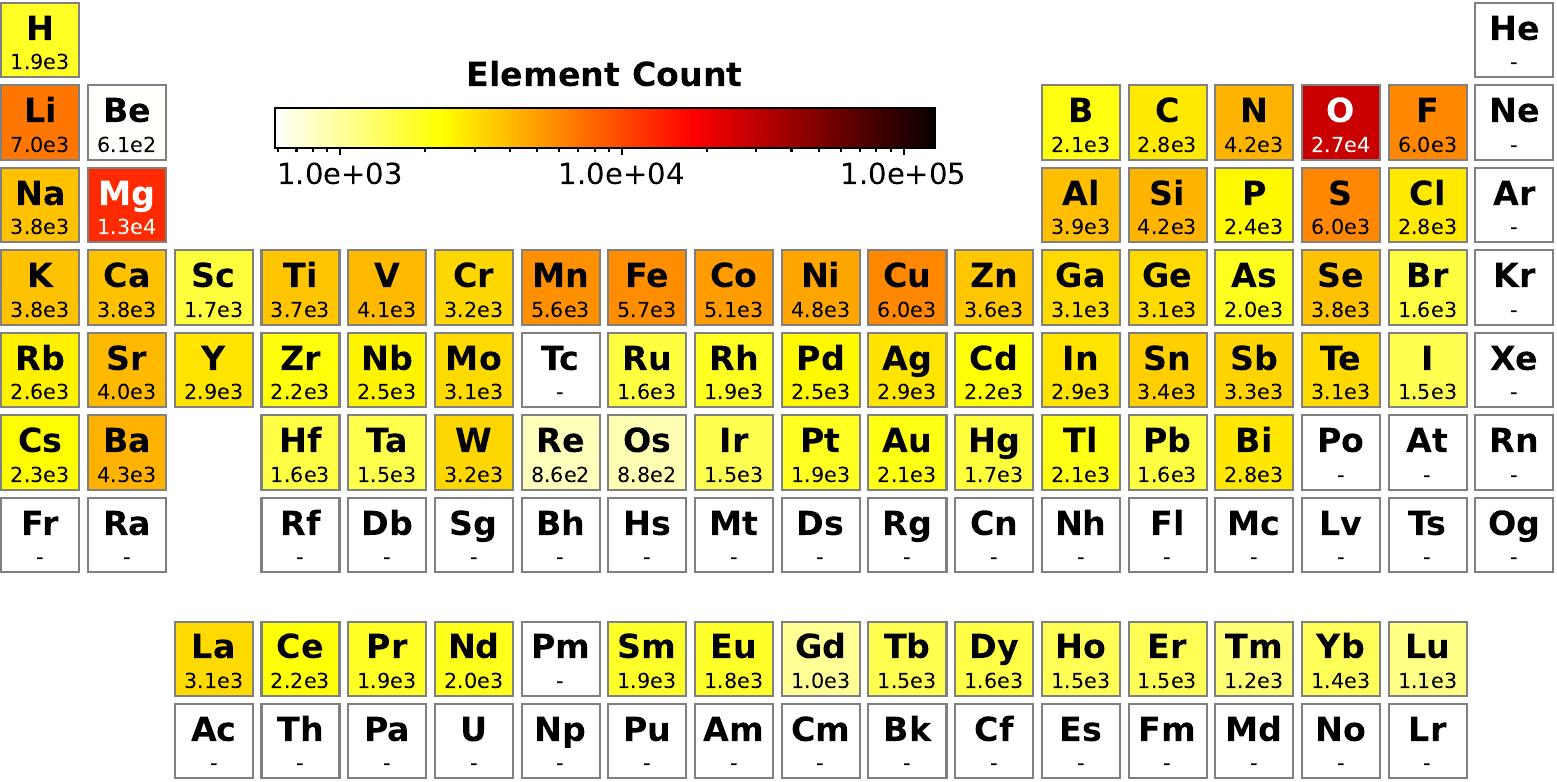}
         \caption{MP}
     \end{subfigure}
     \hfill
     \begin{subfigure}[b]{0.8\textwidth}
         \centering
         \includegraphics[width=\textwidth]{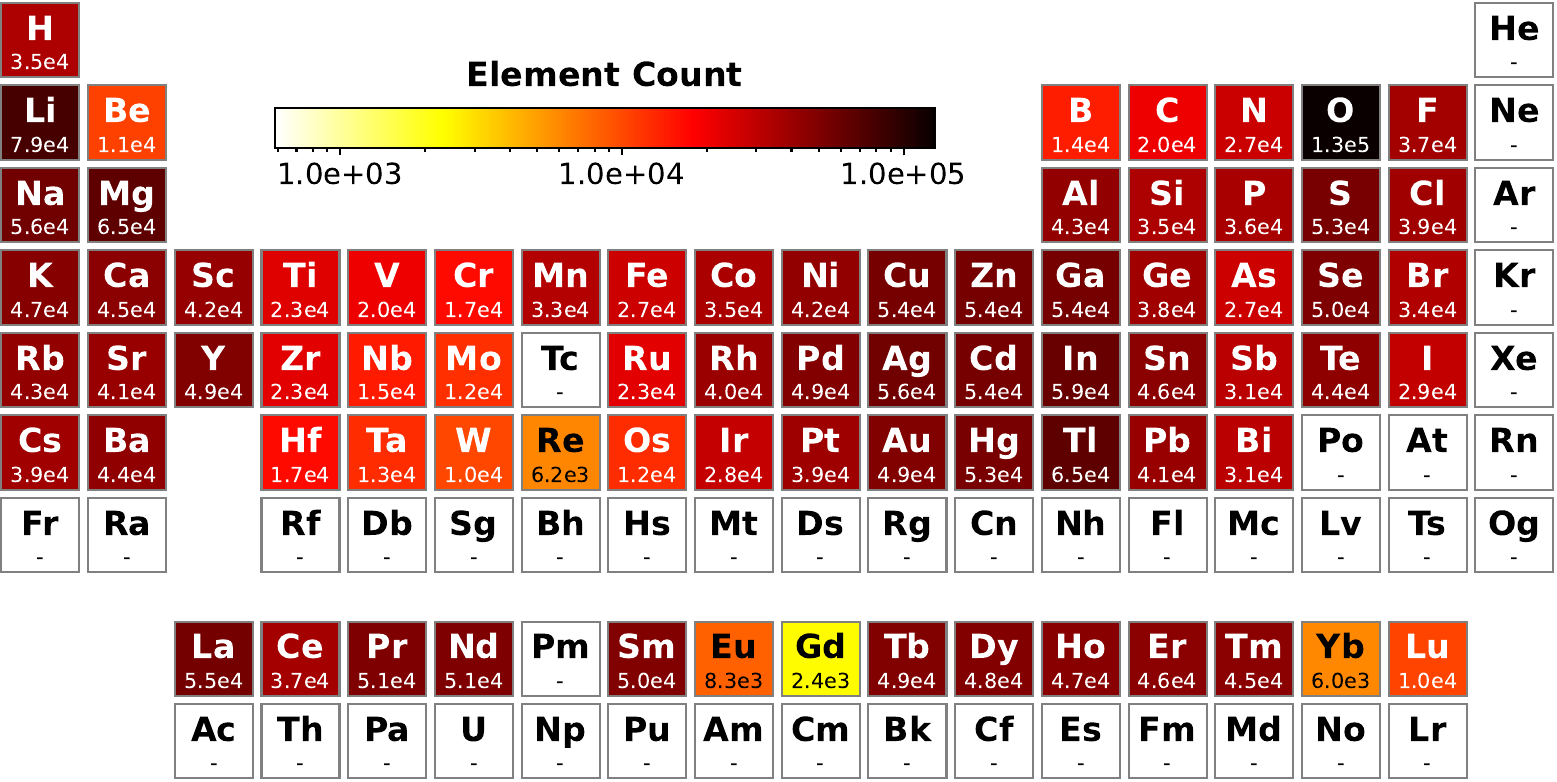}
         \caption{\gls{Alex-MP-ICSD}}
     \end{subfigure}
     \caption{Distribution of elements in \gls{MP} (top) and the combined \gls{Alex-MP-ICSD} (bottom) datasets. Plot generated by pymatviz. \cite{riebesell_pymatviz_2022}}
     \label{fig:elem-dist}
\end{figure}

This process resulted in a reference dataset of 1,081,850 unique structures with associated energy above hull values calculated using \gls{DFT}. We refer to this as the \gls{Alex-MP-ICSD} dataset. This dataset was then used to derive the \gls{Alex-MP-20} dataset, whose element distribution is shown in \cref{fig:elem-dist}.
The \gls{Alex-MP-ICSD} dataset is used as a reference for the computation of stability (i.e., energy above hull) and uniqueness of generated structures.
To train \mbox{MatterGen}, we employ a subset of the \gls{Alex-MP-ICSD} dataset, selecting only structures with up to 20 atoms and whose energy above hull is below 0.1 eV/atom; we refer to this as the \gls{Alex-MP-20} dataset.
Furthermore, we manually exclude from the \gls{Alex-MP-20} dataset all structures belonging to the ``well-explored'' chemical systems, as defined in \cref{sec:supp_chem_sys}.
Additionally, we reserve structures present only in \gls{ICSD} for testing purposes, and therefore exclude them from the \gls{Alex-MP-20} dataset.
We report in \cref{fig:datasets-venn} the structure provenance and quantity for the reference (\gls{Alex-MP-ICSD}) and training (\gls{Alex-MP-20}) datasets.
Finally, the dataset employed to train the \mbox{MatterGen}-MP model contains structures from the \gls{MP}-20 dataset (containing structures with up to 20 atoms) whose energy above hull is below 0.1 eV/atom from the reference convex hull.
This is also highlighted in \cref{fig:datasets-venn} (right panel, blue circle).

\subsection{DFT details}
\label{sec:dft_details}

All \gls{DFT} calculations were performed using the \gls{VASP} \cite{kresse1996efficient} within the projector augmented-wave formalism via \texttt{atomate2}, \texttt{pymatgen} and \texttt{custodian} \citep{ong2013python}. All parameters of the calculations were chosen to be consistent with \gls{MP} \citep{jain2013commentary}, including use of the \gls{PBE} functional \citep{perdew1996generalized} and Hubbard U corrections. Specifically, the following workflows were used for the calculation of associated properties:

\begin{itemize}
    \item Structural relaxations and static calculations for total energy were calculated with the \texttt{MPRelaxSet} settings and via the \texttt{DoubleRelaxMaker} and \texttt{StaticMaker} classes.
    \item Band gaps were calculated as above, but via the \texttt{BandStructureMaker} class.
    \item Elastic tensors were calculated with the default \texttt{ElasticMaker} class with a stencil including -0.01 \% and 0.01 \% deformations, as specified by \texttt{wf\_elastic\_constant\_minimal} in the \texttt{atomate} code. The ``minimal'' preset is used for reasons of computational efficiency.
\end{itemize}

All settings were previously benchmarked or in use by \gls{MP}, and every effort was made to ensure consistent settings were used in the current work.

\newpage

\section{Results}\label{app:exp}
This section contains supplementary information for the results detailed in \cref{sec:results}.

\subsection{Common experimental details}

In this section we provide information about settings used across different experiments, including training and sampling details of \mbox{MatterGen}, and additional information about the \gls{MLFF} we employ. Details specific to certain experiments are deferred to \cref{subsec:unconditional_supp,sec:supp_chem_sys,sec:supp_designing_space_group,sec:supp-single-prop,sec:supp_multi-prop}.

\subsubsection{Hyperparameters for training base model}
The base unconditional model was trained for 1.74 million steps with a batch size of 64 per GPU over eight A100 GPUs using the Adam optimizer \citep{kingma2014adam}. The learning rate was initialized at 0.0001 and was decayed using the ReduceLROnPlateau scheduler with decay factor 0.6, patience 100 and minimum learning rate $10^{-6}$. For the training loss we use $\lambda_\textnormal{coord}=0.1$ and $\lambda_\textnormal{cell}=\lambda_\textnormal{types}=1$, as well as the recommended value $\lambda_\textnormal{CE}=0.01$ for the \gls{D3PM} cross-entropy loss.

\subsubsection{Hyperparameters for fine-tuning models}
\label{sec:supp_hparams_finetune}
For fine-tuning the base model towards different properties, we used a global batch size of 128 and the Adam optimizer. Gradient clipping was applied by value at 0.5. The learning rate was initialized at $6\times 10^{-5}$ and we used the same learning rate scheduler as that for the base model. The training was stopped when the validation loss stopped improving for 100 epochs, which resulted in 32,000 - 1.1 million steps depending on the dataset.

\subsubsection{MatterGen sampling parameters}
\label{sec:mattergen-sampling-params}
For both unconditional and conditional generation, we discretized the reverse diffusion process over the continuous time interval $[0, 1]$ into $T = 1000$ steps. For each time step, we use ancestral sampling (according to \cref{eq:dsm_reverse,eq:ddpm_reverse,eq:d3pm_reverse}) to sample $(\coords\tindex{\tdisc-1}, \lattice\tindex{\tdisc-1}, \types\tindex{\tdisc-1})$ given $(\coords\tindex{\tdisc}, \lattice\tindex{\tdisc}, \types\tindex{\tdisc})$ using the score model described in \cref{app:architecture}. After each predictor step, one corrector step was applied. We used the Langevin corrector (see Algorithms 4 and 5 in \citep{score_sde}) for the coordinates $\coords\tindex{\tdisc}$ and the lattice $\lattice\tindex{\tdisc}$ with signal-to-noise ratio parameters 0.4 and 0.2, respectively. 

\subsubsection{Machine learning force field (MatterSim) details}
\label{sec:mlff_details}

We used an \gls{MLFF} trained on 1.08 million crystalline structures which employed the M3GNet \citep{chen2022universal} architecture with three graph convolution layers and had in total 890,000 parameters. To compute the energy above hull, we used the energy correction scheme compatible with \gls{MP} (i.e., \texttt{MaterialsProject2020Compatibility} from \texttt{pymatgen} \cite{ong2013python}). Further details on the \gls{MLFF} will be provided in a separate publication~\cite{mattersim}.

\subsection{Qualitative analysis of generated structures}\label{app:analysis}

Assessing the quality of generated crystals is difficult. In this work, we have used computational metrics to assess the quality of generated materials. However, additional human review of generated materials can be useful to identify failure modes not captured by these metrics. We perform this human analysis of generated crystals in \cref{sec:structure_analysis_unconditional,sec:supp-single-prop-analysis,subsec:supp_multi-prop_analysis}.

Ultimately, it must at least be possible to create the material in a laboratory setting for any hypothetical material to be of practical use, i.e., the material must be synthesizable. Although it is not practically possible to conclusively determine that a material is synthesizable using theoretical and computational evidence alone, we can use computation as a guide. Throughout this work we use the energy above the convex hull, calculated at 0 K under 0 GPa applied stress, as a signal \cite{bartel2022review} that a material might be stable at ambient temperature and pressure. We acknowledge that there are additional calculations we could perform (such as computing the phonon spectra) that would improve upon the approximations we make in this work, but robustly assessing synthesizability of a hypothetical crystal structure is still an open research question.

Energy above hull alone is not sufficient, since many metastable materials with finite energy above hulls (``off-hull'') are routinely synthesized, while many on-hull materials have not been, despite best efforts. While an attempt \cite{aykol2018thermodynamic} has been made to suggest reasonable threshold values for energy above hull, this is very dependent on chemical system: while some materials (such as carbides, nitrides) are able to tolerate very high energies above convex hull (0.1--1 eV), other materials (such as intermetallics) are only able to tolerate a very small degree of metastability (meV). It is also known \cite{isaacs2018performance} that traditional methods of simulation using \gls{DFT} will give inaccurate energies for some materials, even after empirical corrections \cite{wang2021framework} are applied which can correct some of the better understood systematic errors. Furthermore, the energy above hull is a measure that is only meaningful if the particular chemical system has been well-explored: for unexplored or partially-explored chemical systems, the energy above hull might be inaccurate simply due to more energetically-favourable phases being unknown.

Using empirical priors to assess the synthesizability of structures is traditionally done by hand by domain experts and is therefore highly dependent on the chemical intuition and background knowledge of a particular system by the expert. It becomes more difficult for a scientist to evaluate a crystal structure picked ``at random'', especially as the number of elements increases, as they are less likely to have encountered this material in their prior work. Some materials are hard to analyze in this way because the prior is simply not known or may be biased; for example, any distribution of a specific property calculated from crystalline materials that have already been synthesized will include bias by not simply including the properties of materials that have not yet been synthesized. These biases can be because certain elements are more abundant, cheaper, or easier to process, or because certain materials have garnered more technological interest, rather than because of an \textit{a priori} physical reason why those materials could not be made. Simply put, we do not yet know what the distribution of ``possible'' crystal structures looks like, even within certain constraints (e.g., maximum primitive cell size or number of elements). Therefore, assessing synthesizability using empirical priors is difficult, but might still provide insights. When examining a generated material, in addition to performing a literature review, additional factors can be considered, including:

\begin{enumerate}\label{app:list_empirical_priors}
    \item Symmetry; crystals are defined by their symmetry, and nature prefers symmetrical crystals unless there is a specific mechanism by which symmetry is broken. While lower symmetry would be expected as the number of elements in a system increases, in general generated crystals are expected to be symmetrical. P1 crystals are rare in nature, and when they are reported in databases this is often either because they have not been refined or even because of mis-identification \cite{marsh1999p1}.
    \item The presence of defects. Defects could include structural distortions, with a crystal containing the ``correct'' atoms but in distorted geometry, or could include point defects, such as vacancies or interstitial atoms. The presence of defects is not necessarily bad; for example, many ``off-stochiometric'' materials such as \ce{NiO_x} are routinely synthesized which can contain large concentrations of vacancies. Sometimes vacancies might be required so that a structure might charge balance; a classic example might be the bixbyite crystal structure derived from a fluorite structure with an ordered array of structural vacancies. However, the presence of defects could also be a concern, especially in the case of unexplored chemical systems where a material might be erroneously calculated to be ``on hull'' due to incomplete knowledge of that chemical system. 
    \item Local atomic environments should be reasonable, meaning that the material contains reasonable bond lengths and co-ordination polyhedra, where other materials with those combinations of elements are known.
    \item The material should charge balance if highly ionic, meaning that the sum of the formal valence of its constituent atoms is zero. If a material does not charge balance, and is ionic, it will likely have a very low defect formation energy if synthesizable. However, the importance of charge balance should be taken with caution, since the proportion of new materials that have been discovered that are nominally charge-balanced has decreased over time \cite{antoniuk2023predicting}, with only around 40\% newly-discovered materials being charge-balanced compared to over 80\% of materials discovered a hundred years ago.
\end{enumerate}

The definition of a material used in this work (\cref{sec:results}) also allows for many potential failure modes including but not limited to: non 3D periodic materials such as 2D materials that contain a vacuum in one dimension, amorphous materials, or other random arrangements of ``atoms-in-a-box''. Effort can be made to avoid these classes of materials by altering the training data of the generative model. Some efforts have been made to algorithmically categorize crystals \cite{himanen2018materials}, which could then be used to this effect, however these tools are not yet well-developed, and some spurious structures in the training set should be expected.
Finally, the definition of a material used in this work also explicitly assumes an ordered material, whereas many real materials exist as alloys or solid solutions: in traditional \gls{DFT}, only fully ordered materials can be calculated (meaning, a material that contains whole atoms, with exactly one atom on a given atomic site). This is also the constraint placed by \mbox{MatterGen} on its generated crystals. However, real materials are often disordered, with fractional atomic occupancy (on \emph{average}, a given atomic site might contain, say, 50\% of one atom, and 50\% of another atom). A disordered material has many ``ordered approximations''--small unit cells with the correct overall composition--that can represent the parent disordered material. As such, \mbox{MatterGen} might generate several ordered approximations of the same parent disordered material. In these cases, novelty will not be assessed correctly, and properties such as energy above hull might be misleading.

Given these factors, we acknowledge that there are limitations in materials discovery efforts that still require methods advancements to overcome. We restrict our confidence that structures we generate are synthesizable to the level of theory and computation we use in this work, in addition to the finite reference we use for the convex hull. We have attempted human-assisted evaluation of predicted structures using empirical priors to gain additional knowledge for how our model performs at generating synthesizable materials.

\subsubsection{Visualization}

Manual analysis of crystal structures can be influenced by how they are represented visually, for example the specific bonds that are drawn. In this work, crystal structures are visualized using Crystal Toolkit \cite{horton2023crystal} with the CrystalNN \cite{pan2021benchmarking} bonding algorithm, since this is known to give good results in most cases. A uniform atomic radius was used since a wide variety of chemical bonding is expected to be present, and no one type of atomic radii (covalent, ionic, etc.) can be assumed. When valences are indicated, these are formal valences assigned using heuristic methods in \texttt{pymatgen} \cite{ong2013python}. With the exception of \cref{fig:illustrative}, all visualizations are of a $2 \times 2 \times 2$ supercell to ensure at least one full repeat of a crystal and its periodic images are shown. All visualizations are of crystal structures as-calculated, meaning they are not necessarily in their conventional setting, and therefore axes are not labelled.

\subsection{Generating stable and diverse materials}
\label{subsec:unconditional_supp}

This section provides supplementary information to the results in \cref{sec:generating_stable_diverse}.

\subsubsection{RMSD, stability, uniqueness and novelty}
\label{sec:rmsd-stability-uniqueness-novelty}
To evaluate the performance of a generative model on the task of unconditional generation, we look at two keys metrics. First, we use the \gls{RMSD} between the generated and the \gls{DFT}-relaxed structure to measure local stability. Second, we use the fraction of \gls{SUN} structures to capture global stability and, to some extent, diversity.
The \gls{RMSD} metric is defined as
\begin{equation}
    RMSD = \sqrt{\min_\mP \dfrac{1}{N} \sum_n^N \left|\cartcoord_{\mP(n)}^{gen} - \cartcoord^{DFT}_n\right|^2},
\end{equation}
where $\cartcoord_n$ indicates the Cartesian coordinates of atom $n$, and $\mP$ is the element-aware permutation operator on the atoms of the generated structure.
A lower \gls{RMSD} indicates that generated structures are closer to their \gls{DFT}-relaxed counterpart. This in turn reduces the computational time for the \gls{DFT} relaxation, which is typically the most costly part of crystal structure generation.
Novelty and uniqueness are computed in all model evaluations by comparing the atomic arrangement of every pair of structures that have the same reduced formula and space group via the \texttt{StructureMatcher} utility from the \texttt{pymatgen} Python package \cite{ong2013python}, with the following default parameters: \texttt{ltol=0.2}, \texttt{stol=0.3}, \texttt{angle\_tol=5}. This definition of novelty is not able to determine whether an ordered structure might be an ordered approximation of a disordered structure, and so some structures might be falsely determined to be novel in this scenario.

The plots displayed in \cref{fig:unconditional}(b,c,e,f), and the structures showed in \cref{fig:unconditional}(1) refer to samples of 1024 generated structures which have also been relaxed via \gls{DFT}. 

\subsubsection{Additional qualitative analysis of structures} \label{sec:structure_analysis_unconditional}
Within the 1024 structures generated unconditionally for \cref{fig:unconditional}(a), a total of 43 unique, on-hull crystal structures were found: 11 binaries, 22 ternaries, and 10 quaternaries. These are summarized in \cref{tab:unconditional_analysis}. Of these, 3 had P1 symmetry, and 3 contained molecules or were molecular crystals. Prototype assignment limited to prototypes available in the \texttt{robocrystallographer} \cite{ganose2019robocrystallographer} tool, which will assign the ``closest'' matching prototype subject to tolerances. The ability for a composition to charge balance assessed by \texttt{pymatgen} and known common oxidation states for each element. As previously discussed, a generated crystal can still be reasonable even if it does not charge balance, and not all materials are ionic.

Four randomly-selected examples were highlighted in \cref{fig:unconditional} in the main text. These were \ce{BaLa2Ir}, \ce{K3AlCl6}, \ce{NaNiIO6}, and \ce{NaSmTm2Te4}. For \ce{BaLa2Ir}, it is well-known that \ce{La2Ir} forms an intermetallic with a Laves structure, and that Ba can often substitute for La since both can exist in a +2 oxidation state with a 6s2 outer shell. However, in this example, we see octahedrally co-ordinated Ir bonded to La, with Ba inserted as a single plane of atoms in a close-packed configuration, as it would exist in elemental Ba. It is unclear if this material could exist. The \ce{K3AlCl6} has low space group symmetry (P$\overline{1}$), but consists of Al in an ideal octahedral co-ordination, with K in a mixed bonding environment. This structure charge balances under the assumption of \ce{Al^{3+}}, \ce{K+} and \ce{Cl-}, and could be thought of as derived from a defected rocksalt. The \ce{NaNiIO6} structure exists in the training set and is well-known belonging to a family of periodate structures \ce{AMIO6} (where A is an alkali metal and M is another metal atom). This material is therefore an example of a material incorrectly classified by our novelty filter; a material might be classified as novel prior to DFT relaxation, and not after relaxation. \ce{NaSmTm2Te4} is a rocksalt structure with Te in the anion site and a mix of Na, Sm and Tm in the cation sites.

\begin{table}[th]
    \centering
\resizebox{\columnwidth}{!}{%
\begin{tabular}{llrlll}
\toprule
           Formula &          Symmetry &  \# Elements & Charge balances &                    Prototype &         Classification \\
\midrule
        \ce{Ba4Au} &              C2/m &           2 &               - &                            - & Crystal with 1D chains \\
         \ce{SbAu} &      P6$_{3}$/mmc &           2 &             Yes & Molybdenum Carbide MAX Phase &           Bulk crystal \\
         \ce{ReF6} & Im$\overline{3}$m &           2 &             Yes &                     Tungsten &      Molecular crystal \\
      \ce{Tb2Zn17} &  R$\overline{3}$m &           2 &               - &                            - &           Bulk crystal \\
          \ce{SeS} &                P1 &           2 &             Yes &                 red selenium & Crystal with 1D chains \\
        \ce{V3Cl8} &              C2/m &           2 &             Yes &                            - &     Layered/2D crystal \\
         \ce{KCl8} &                P1 &           2 &               - &                            - &                 Hybrid \\
         \ce{AlI7} &                P1 &           2 &               - &                            - & Crystal with 1D chains \\
         \ce{VBr5} &                Cm &           2 &             Yes &        Silicon tetrafluoride &      Molecular crystal \\
        \ce{Li4Hg} &              I4/m &           2 &               - &                            - &           Bulk crystal \\
        \ce{DyIn3} & Pm$\overline{3}$m &           2 &               - &             Uranium Silicide &           Bulk crystal \\
       \ce{HfAlAu} & P$\overline{6}$2m &           3 &               - &                            - &           Bulk crystal \\
      \ce{Lu2AgOs} &            P4/mmm &           3 &               - &                      Heusler &           Bulk crystal \\
       \ce{GdScBi} &            P4/nmm &           3 &               - &                   Matlockite &           Bulk crystal \\
     \ce{Sm(FeC)2} &              Fddd &           3 &             Yes &                            - &           Bulk crystal \\
    \ce{Li(AlPd)2} &            P4/mbm &           3 &               - &                            - &           Bulk crystal \\
      \ce{YbNiSn2} &              Cmcm &           3 &               - &                            - &           Bulk crystal \\
    \ce{Nd(GaPt)2} &            P4/nmm &           3 &               - &                            - &           Bulk crystal \\
       \ce{LiPrAs} & P$\overline{6}$m2 &           3 &               - &                            - &           Bulk crystal \\
    \ce{Eu(AgSe)2} & P$\overline{3}$m1 &           3 &             Yes &                            - &           Bulk crystal \\
      \ce{Tl4IrO6} &              C2/m &           3 &             Yes &                            - &           Bulk crystal \\
      \ce{CsTe2Pd} &              C2/m &           3 &             Yes &                            - &           Bulk crystal \\
      \ce{Al2PdI8} &              C2/m &           3 &             Yes &                       Indium &      Molecular crystal \\
      \ce{CaTbCd2} &            P4/mmm &           3 &               - &                      Heusler &           Bulk crystal \\
      \ce{Hf2ZnMo} &            P4/mmm &           3 &               - &                            - &           Bulk crystal \\
      \ce{Sb5PPb2} &              Amm2 &           3 &             Yes &                            - &           Bulk crystal \\
     \ce{La6SbTe5} &              C2/m &           3 &             Yes &             Caswellsilverite &           Bulk crystal \\
       \ce{CeTeAs} &              Pnma &           3 &             Yes &                            - &           Bulk crystal \\
       \ce{NaH3Pd} & Pm$\overline{3}$m &           3 &             Yes &           (Cubic) Perovskite &           Bulk crystal \\
     \ce{Er2ZnNi2} &              Immm &           3 &               - &                            - &           Bulk crystal \\
        \ce{YBeSi} &      P6$_{3}$/mmc &           3 &               - &                            - &           Bulk crystal \\
     \ce{TePb5Cl8} &              C2/m &           3 &             Yes &                            - &           Bulk crystal \\
       \ce{FeCoH2} &            P4/mmm &           3 &               - &             Caswellsilverite &           Bulk crystal \\
   \ce{Sc4GaCu2Rh} &  R$\overline{3}$m &           4 &               - &                      Heusler &           Bulk crystal \\
  \ce{La8Os3PdBr4} &  R$\overline{3}$m &           4 &               - &             Caswellsilverite &           Bulk crystal \\
\ce{TbCe(Ho2Te3)2} &                Cm &           4 &               - &                     alpha Po &           Bulk crystal \\
   \ce{Ba8Sb3PBr4} &  R$\overline{3}$m &           4 &             Yes &                     alpha Po &           Bulk crystal \\
     \ce{Cs2KZnF6} & Fm$\overline{3}$m &           4 &               - &           (Cubic) Perovskite &           Bulk crystal \\
     \ce{Zn2Ni6BH} & Fm$\overline{3}$m &           4 &               - &           (Cubic) Perovskite &           Bulk crystal \\
    \ce{Cs2AuIBr6} &            I4/mmm &           4 &             Yes &                            - &           Bulk crystal \\
\ce{LiTb3(DySe3)2} &              C2/m &           4 &               - &                     alpha Po &           Bulk crystal \\
   \ce{Ba8Pd2RhAu} &              P2/m &           4 &               - &        beta Vanadium nitride &           Bulk crystal \\
    \ce{Te2MoWSe2} &                Cm &           4 &             Yes &                  Molybdenite &     Layered/2D crystal \\
\bottomrule
\end{tabular}
}

\caption{Summary information on 43 crystal structures that were calculated as being thermodynamically stable from a batch of 1024 crystal structures from an unconditional generation task.}
\label{tab:unconditional_analysis}
\end{table}

\subsection{Generating materials with target chemistry}
\label{sec:supp_chem_sys}

This section provides supplementary information to the results in \cref{sec:chemical-system}.

\subsubsection{Additional experimental details}
We explore the capability of \mbox{MatterGen} to find novel stable crystals across the 27 chemical systems listed in \cref{tab:chemical_systems_description}.
We group the systems in terms of how many elements they contain (ternary, quaternary, and quinary), and in terms of how many structures near the the convex hull were present in the reference \gls{Alex-MP-ICSD} dataset (`well explored', `partially explored', `not explored').
The latter classes are defined as follows:
\begin{itemize}
    \item `well explored': systems with the highest numbers of structures near the convex hull. We removed all structures belonging to `well explored' systems from the training data set to assess the capability of our model to recover existing stable structures without having seen them during training.
    \item `partially explored`: systems that lie between the 30th and the 90th percentile of the distribution of chemical systems based on the number of structures they have near the convex hull. This class was designed to assess the capability of our model to expand known convex hulls. Therefore, we did not remove the existing data belonging to such systems from the training set.
    \item `not explored': systems with no data near the convex hull. This class was designed to test our model in chemical systems where no structures on the hull are present in the reference dataset.
\end{itemize}
Here, we define `near the convex hull' structures as structures whose energy per atom is between 0.0 and 0.1 eV/atom above the convex hull.
For all three groups, we randomly chose nine ternary, nine quaternary and nine quinary chemical systems (see \cref{tab:chemical_systems_description}). Moreover,  we replaced those chemical systems that had an overlap of more than two elements with another system to promote chemical diversity. The replacement was chosen randomly as well.

For this task, we fine-tune our base model on two properties: chemical system and energy above hull.
We encode the latent embedding for the energy above hull and the chemical system as detailed in \cref{subsec:continuous_embedding} and \cref{subsec:discrete_embedding}, respectively.
Both properties are available for all structures in the training set of the base model. Therefore, the training set is used in full for fine-tuning.
At sampling time, we condition on both an energy above the convex hull of $0.0$~eV/atom, and on the chemical system we want to sample.

To compare the performance of MatterGen against substitution and \gls{RSS}, we employ an \gls{MLFF} (\mbox{MatterSim}, see \cref{sec:mlff_details}) to relax the generated structures, and then perform \emph{ab initio} relaxation and static calculations via \gls{DFT} (see \cref{sec:dft_details} for details).
In particular, we perform the following steps: (1) generate structures, (2) relax structures using the \gls{MLFF}, (3) filter structures for uniqueness, (4) select the 100 structures with lowest predicted energy above hull according to the \gls{MLFF}, (5) run \gls{DFT} on these structures. We report metrics only with respect to those structures.
To allow for a fair comparison between our generative model and non-generative approaches, we employ the \gls{MLFF} relaxation on a greater number of samples for the latter.
For \gls{RSS}, we sample 600,000 structures per chemical system according to the protocol described in \cref{sec:rss_details}.
For substitution, we enumerate every possible structure according to the algorithm detailed in \cref{sec:substitution_details}, which yields between 15,000 and 70,000 structures per chemical system.
For MatterGen, we generate 10,240 structures per chemical system.

\begin{table}[ht!]
    \centering
    \begin{tabular}{lccc}\toprule
         & \textbf{Ternary} & \textbf{Quaternary} & \textbf{Quinary}\\\midrule
         & O-Sr-V & Bi-Cu-Pb-S & C-H-N-O-S \\
\textbf{Well explored} & Mn-O-Se & As-Cl-O-Pb & Ba-Ca-Cu-O-Tl \\ 
         & La-Mo-O &Fe-Na-O-P & Eu-F-K-O-Si \\ 
         \hline
         & Li-Pr-Te & Cl-Cu-Dy-Rb & Cu-Gd-O-Ru-Sr \\
\textbf{Partially explored} & C-Pr-Ru & Na-Te-Tm-Zr & La-Na-O-Sb-Sc \\ 
         & C-Mg-Sc & F-Mg-Rb-Sn & Cs-F-O-Tl-Zr \\ 
        \hline
         & Br-Pb-Rh & Cr-Ga-Mg-P & Al-C-H-Sb-Zr \\
\textbf{Not explored} & As-Cu-Sr & C-Cl-Ho-Ru & As-Br-Cr-I-Pt \\ 
         & Cl-Er-In & Al-Au-Co-S & K-Mo-O-P-Sr \\\bottomrule
    \end{tabular}
    \caption{Categorization of the 27 chemical systems used to benchmark model capabilities on chemical system exploration}
    \label{tab:chemical_systems_description}
\end{table}

\paragraph{Random structure search details}
\label{sec:rss_details}
For every chemical system, we performed two rounds of \gls{RSS} using the \texttt{airss} \cite{pickard2011ab} package. In each round, we generated 300,000 structures by sampling 100,000 structures across three different ranges of number of atoms per unit cell. We used the following non-overlapping intervals: 3-9, 10-15, and 16-20 for ternary systems; 4-10, 11-15, and 16-20 for quaternary systems; 5-11, 12-16, and 17-20 for quinary systems. For the first round, we used \texttt{airss} to propose structures without structural relaxation using \texttt{\mbox{MINSEP = 0.7-3}} (minimum separation between atoms in \si{\angstrom}) and \texttt{\mbox{SYMMOPS = 2-4}} (number of symmetry operations). 
All proposed structures were relaxed using an \gls{MLFF} (\mbox{MatterSim}, see \cref{sec:mlff_details}), and the resulting 300,000 \gls{MLFF} relaxation trajectories were used in the second round of \gls{RSS} to automatically tune the \texttt{MINSEP} parameter. Again, \texttt{airss} was run without structural relaxation followed by a \gls{MLFF} relaxation. Finally, we combined the 600,000 \gls{MLFF}-relaxed structures from both rounds and ran \gls{DFT} structural relaxation and static calculation on the 100 unique structures with the lowest predicted energy above hull according to the \gls{MLFF}.

\paragraph{Substitution details}
\label{sec:substitution_details}
A total of 5,143 ordered crystal structures (2,695 ternary, 1,875 quaternary, and 573 quinary) with less than 100 atoms in a unit cell from the \gls{ICSD} \cite{zagorac2019recent} were used as prototypes. For each chemical system in \cref{tab:chemical_systems_description}, we computed all possible unique substitutions of the prototypes, relaxed all structures using a \gls{MLFF} (\mbox{MatterSim}, see \cref{sec:mlff_details}), and selected the 100 unique structures with the lowest predicted energy above the hull according to the \gls{MLFF}. Finally, we ran \gls{DFT} structural relaxation and static calculation on the selected structures.

\subsubsection{Additional qualitative analysis of structures} \label{subsec:structure_analysis_chemical_system}

The V-Sr-O chemical system example provided in \cref{fig:chemsys} produced four new on-hull crystal structures: \ce{SrV2O6} (\ce{V^{5+}}), \ce{SrVO3} (\ce{V^{4+}}), \ce{Sr3V2O8} (\ce{V^{5+}}) and \ce{SrV2O4} (\ce{V^{4+}}). This chemical system has been well-studied in literature, with \ce{SrVO3} being a well-known perovskite \cite{nekrasov2005comparative}, \ce{Sr2VO4} expected to crystallize into a \ce{K2NiF}-like crystal structure, and \ce{Sr3(VO4)2} synthesized in a cation-deficient variant of the \ce{SrVO3} crystal structure \cite{pati2013phase}. 

Vanadates are known to by synthesizable in a variety of frameworks, with the expected co-ordination of the \ce{VO4} sub-unit varying with oxidation state \cite{zavalij1999structural} from the ideal tetrahedron in \ce{V^{5+}} to a variety of other co-ordination environments. All generated structures have plausible atomic environments with \ce{VO4} sub-units, either ideal or distorted, and oxygen co-ordinated Sr atoms.

One \ce{SrV2O4} structure, having P$\bar{1}$ symmetry, consists of a layers of ideal \ce{VO4} edge-sharing tetrahedra separated by a Sr in a triangular prismatic bonding configuration, resulting in a 1D channel of voids in the Sr layer. 

\subsection{Designing materials with target symmetry}
\label{sec:supp_designing_space_group}

This section provides supplementary information to the results in \cref{sec:designing_space_group}.

\subsubsection{Additional experimental details}
\label{subsec:space_group_supp}
For generating structures belonging to a target space group, we fine-tune our base model on the whole training set, and represent the latent embedding of the space group of a crystal via one-hot encoding of the space group.

\begin{figure}[ht]
    \centering
\includegraphics[width=1\linewidth]{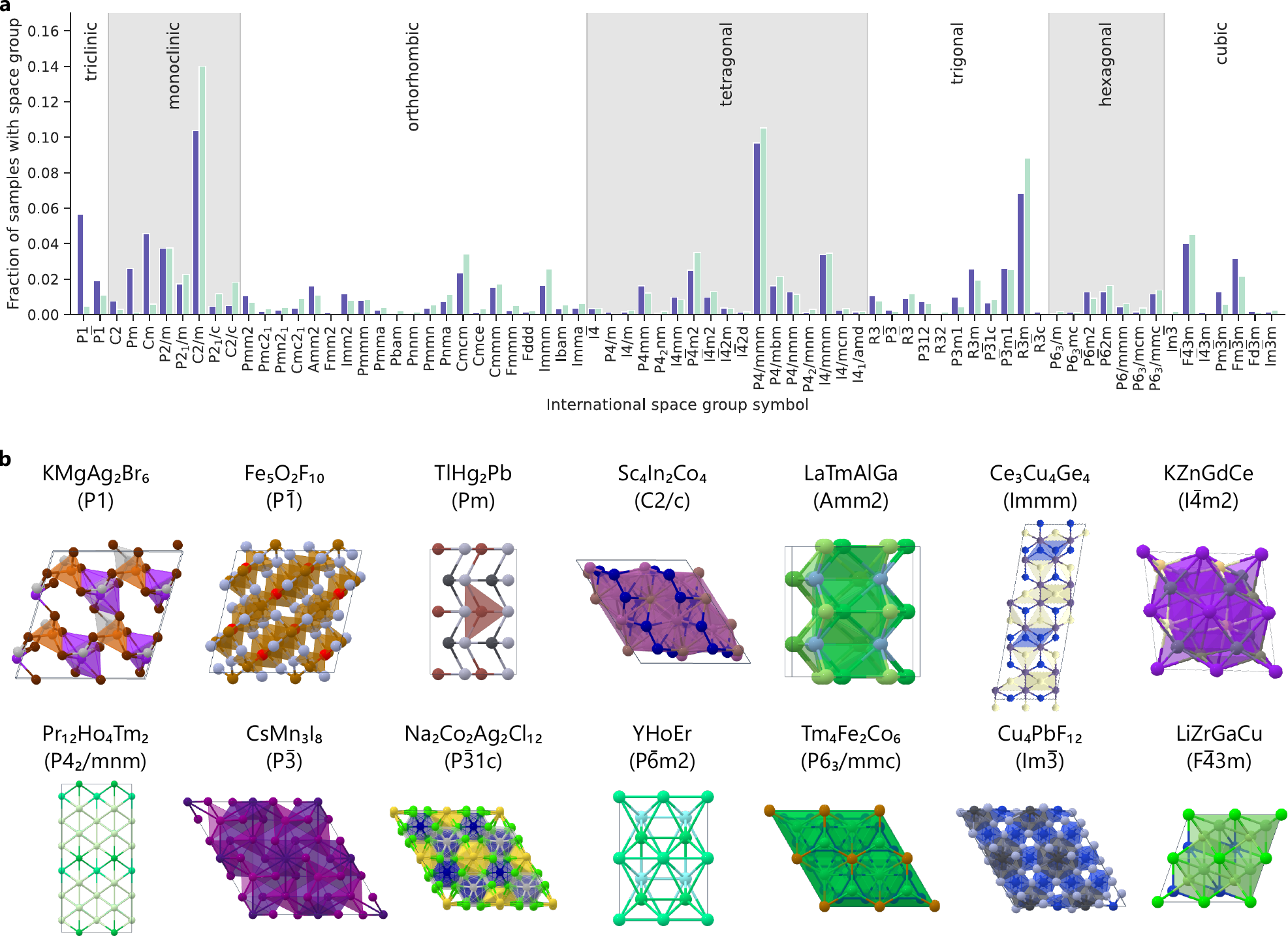}
    \caption{\textbf{Generating materials with target symmetry.}
    (a) Fraction of generated structures belonging to space groups generated by \mbox{MatterGen} fine-tuned on symmetry groups, sampled unconditionally (violet), and structures present in the reference dataset (green).
    (b) Fourteen generated \gls{SUN} structures, one for each space group reported in \cref{fig:symmetry}; composition and space group are reported.
    }
    \label{fig:symmetry_supp}
\end{figure}

We assess the capability of our model to correctly generate structures belonging to any space group via two tasks.
For the first task (\cref{fig:symmetry}), we sample two space groups for each of the seven lattice systems, and choose to sample only from space groups that contain at least 1000 structures in the training set.
We then compute the fraction of \gls{SUN} structures our fine-tuned model generates when conditioned on these space groups that are classified as belonging to that space group according to the \texttt{SpaceGroupAnalyzer} module of \texttt{pymatgen} \cite{grosse2022algorithms, ong2013python}.
This metric is computed for 256 generated structures per space group after \gls{DFT} relaxation has been performed. 
For the second task (\cref{fig:symmetry_supp}), we generate 10,000 structures conditioned on space groups sampled randomly from the data distribution of the training set, and check whether our model is able to reproduce the distribution of space groups from the training data.
For both of the above tasks, the number of atoms in the systems are sampled from the distribution of number of atoms for that space group in the training set.
This way, we avoid `impossible' space group constraints, where the space group we condition on cannot be satisfied given the number of atoms we set.

\subsubsection{Additional Analysis of Structures}
\label{sec:analysis_symmetry}

The highlighted examples in \cref{fig:symmetry} show a high degree of novelty, with few matching known prototypes. The combination of elements in generated structures is also not common, e.g., \ce{DyScNiPd}, \ce{CeAsRh}. It is therefore difficult to reference to known structures in literature. Half of the examples could be assigned formal valences, and the majority were robust for their target symmetry even when symmetry was calculated with a higher symmetry-finding tolerance.

\subsection{Designing materials with target magnetic, electronic and mechanical properties}
\label{sec:supp-single-prop}

This section provides supplementary information to the results in \cref{sec:single-prop}.

\subsubsection{Additional experimental details}

To generate structures conditioned on a target property, we fine-tune our base model on magnetic density ($N\approx$~605,000 \gls{DFT} labels), band gap ($N\approx$~42,000) and bulk modulus ($N\approx$~5,000), respectively. See \cref{sec:fine_tuning} for more details on the fine-tuning scheme, and \cref{sec:supp_hparams_finetune} for hyperparameter settings. We represent the latent embedding of scalar properties via a sinusoidal encoding from the Transformer architecture \cite{vaswani2017attention}.

For each property in \cref{fig:single-prop}(a-c), we generate $512$ samples with our fine-tuned model by conditioning on a value of $0.2~\text{\AA}^{-3}$ for magnetic density, $3.0$ eV for band gap, and $400$ GPa for bulk modulus. We relax those structures using the \gls{MLFF} and filter the relaxed structures by stability  and uniqueness. We then relax the remaining structures with \gls{DFT} and filter by stability, uniqueness, and novelty. Finally, we compute the desired property of the remaining structures using \gls{DFT}, and filter out structures where we consider the computed property to be an outlier, i.e., high bulk modulus values with more than 600 GPa and magnetic density values with more than 0.3$~\text{\AA}^{-3}$. For more details on the \gls{DFT} calculations, see \cref{sec:dft_details}. After all filters and \gls{DFT} computations, we obtain 251 \gls{SUN} structures for magnetic density, 142 for band gap, and 22 for bulk modulus.

For \cref{fig:single-prop}(g), we use \mbox{MatterGen} to generate 15,360 samples conditioned on a magnetic density value of $0.2~\text{\AA}^{-3}$, relax those structures using the \gls{MLFF} and filter the relaxed structures by stability and uniqueness. This results in 5,365 candidate structures. Next, we randomly sub-sample 600 structures and relax them via \gls{DFT}, with 540 of them being \gls{DFT} stable. We then randomly sub-sample $k$ structures according to the given \gls{DFT} property calculation budget, and compute their magnetic density via \gls{DFT}. As a baseline, we count the number of structures in the labeled training dataset that satisfy the target property constraint. For \cref{fig:single-prop}(h), we train a separate property predictor for bulk modulus (see details below), which we use for both \mbox{MatterGen} and the screening baseline. In particular, we use predicted bulk modulus values to fine-tune the base \mbox{MatterGen} model. We generate 8,192 samples conditioned on a bulk modulus value of 400 GPa, relax those structures using the \gls{MLFF} and filter the relaxed structures by stability and uniqueness. This results in 801 candidate structures. We relax those via \gls{DFT}, with 736 of them being \gls{DFT} stable. We then randomly sub-sample $k$ structures according to the given \gls{DFT} property calculation budget, and compute their bulk modulus via \gls{DFT}. For the screening baseline, we use the bulk modulus property predictor to predict the bulk modulus values of all structures in the training dataset for which we do not have an existing \gls{DFT} label, rank those structures by their predicted bulk modulus values, and choose the top $k$ structures according the \gls{DFT} property calculation budget. We then verify their actual bulk modulus values via \gls{DFT}.
 
\paragraph*{Property predictor details}
\label{sec:screening_details}

The bulk modulus property predictor used in \cref{fig:single-prop}(h) consists of a GemNet-dT \cite{gasteiger2021gemnet} encoder that provides atom and edge embeddings, followed by a mean readout layer. We employ three message-passing layers, a cutoff radius of 10~$\text{\AA}$ for the neighbor list construction, and set the dimension of nodes and edges hidden representations to 128. 

To train the property predictor, we use all materials with \gls{DFT} Voigt-Reuss-Hill average bulk modulus values from \gls{MP} \cite{jain2013commentary} (including structures with more than 20 atoms), which are 7,108 structures in total. We allocate $80\%$ of the data for the training set, $10\%$ for validation, and $10\%$ for testing. We follow the MatBench benchmark \cite{dunn2020benchmarking} and predict the $\log_{10}$ bulk modulus. At the end of training, the model achieves a \gls{MAE} of \SI{9.5}{GPa}.
The model was trained using the Adam optimizer. Gradient clipping was applied by value at $0.$5. The learning rate was initialized at $5\times 10^{-4}$ and decayed using the ReduceLROnPlateau scheduler with decay factor 0.8, patience 10 and minimum learning rate $10^{-8}$. The training was stopped when the validation loss stopped improving for 150 epochs.

\subsubsection{Additional qualitative analysis of structures}
\label{sec:supp-single-prop-analysis}

\paragraph*{Magnetic density conditional generation} \label{subsec:structure_analysis_mag_density}
For the high magnetic density generation task, a manual review of ten generated crystals chosen at random was performed, as well as the two representative structures with high magnetization density that were highlighted in \cref{fig:single-prop}. Of the random selection, eight of the ten generated structures were ordered approximations of a Fe-Co alloy: $\alpha$-Fe with a partial Co substitution from 10\% Co to 40\% Co. The Fe-Co system is a well-known and versatile soft magnetic material system with a wide region where a $\alpha$-\ce{Fe_xCo_1-x} phase is stable, with a mixture of $\gamma$-Fe and $\alpha$-Co expected at lower Co content. While these generated structures could be considered ``good'', in the sense that they are physically plausible and indeed would be useful magnetic materials, they are likely only considered novel since these specific ordered unit cells do not currently exist in the reference databases. Also, as an alloy, predicted quantities such as energy above hull based on a single ordered approximation will be misleading.
Two of the ten generated structures contained hydrogen, either as \ce{Fe9H2} or \ce{Fe4Co2H}, with hydrogen in an ideal octahedral environment. The generation of these structures can be rationalized owing to extensive study of iron hydrides, with the generated structures having local atomic environments similar to that in a double hexagonal close-packed iron hydride structure that can exist at high pressure. The realization of these specific structures is unlikely with phase segregation expected to occur. The two structures highlighted in \cref{fig:single-prop}(d) contain Gd, an element with a large magnetic moment due to its 7 unpaired electrons when in its \ce{Gd^{3+}} oxidation state and in its elemental state. Therefore, it is unsurprising that the model would preferentially generate materials containing Gd in this instance: as in the randomly selected sample, this task necessarily shows a strong compositional bias in the materials generated. The \ce{Gd2N} example is a layered material. Although the \ce{Gd2N} molecule has been studied \cite{willson1998characterization}, it is unknown if and how it might crystallize. The generated \ce{Gd6H2CN3} structure is rocksalt-derived, with Gd on one site, and a mix of C, H and N on another site, with all atoms in almost ideal octahedral environments.


\paragraph{Band gap conditional generation} \label{subsec:structure_analysis_band_gap}
For the target band gap generation task, a manual review of ten generated crystals chosen at random was performed, as well as the two representative structures with desired target band gap that were highlighted in \cref{fig:single-prop}. Unlike the other single property optimization tasks, the generated crystals did not show a strong compositional dependence, with a wide range of elements presented in generated materials. All generated materials also could nominally charge balance, unlike the materials analyzed from other single property optimization tasks. This seems reasonable for a task designed to generate insulating systems; the target band gap of 3 eV (calculated with the PBE functional, and thus underestimating the true electronic band gap) should generate insulating materials, and thus more ionic solids. Of the random sample, we could only find \ce{NaNO3} and the molecular crystal \ce{BI3} as a compositions that had previously been synthesized. Both were in incorrect symmetry compared to their experimentally known structures, with the generated \ce{NaNO3} in C2/c compared to the experimental R$\overline{3}$c, and \ce{BI3} in P1 compared to the experimental P$6_3$/m. However, both had correct local bonding and similar calculated band gaps to the band gaps calculated for their experimental structures. This indicates one instance whereby the model might still generate useful results even if the generated crystal structure is different from the experimental ground state, if it can guide a scientist towards investigation of a specific system.

The examples highlighted in \cref{fig:single-prop}(e) were \ce{VBiO4} and \ce{TlNO3}. Of these, the local bonding in \ce{VBiO4} was very similar to that in the experimentally-known bismuth vanadate which crystallizes in the I$4_{1}$/amd \cite{dreyer1981dreyerit} space group or I$4_{1}$/a space group \cite{sleight1979crystal}, unlike the generated P$2_{1}$/m space group, and so is another example of reasonable local environment but with a seemingly incorrect space group. Nevertheless, this crystal structure was calculated to be thermodynamically stable with respect to these experimentally-known crystal structures; this could be a limitation with respect to the DFT methods used, or a sign of under-convergence for several very similar-in-energy polymorphs. As such, it should not be seen as a failure of the model. Likewise, \ce{TlNO3} thallous nitrate is experimentally-known in the Pnma space group \cite{fraser1975nitrate} with a calculated (\gls{PBE}) band gap of 2.8 eV \cite{mp-5915}, however the generated crystal appears quite different to the experimental structure due to a change in Tl co-ordination around \ce{NO3^-} anions.

\paragraph{Bulk modulus conditional generation}\label{subsec:structure_analysis_bulk_mod}
For the high bulk modulus generation task, a manual review of ten generated crystals chosen at random was performed, as well as the two representative structures with high bulk modulus that were highlighted in \cref{fig:single-prop}(f). All structures show a strong compositional bias, containing a mix of refactory elements Re, W, Mo and Ir and frequently also B and C; this is consistent with literature on superhard materials (which includes materials with high bulk modulus). When the composition contains only refactory elements, the generated structure typically seems like an ordered approximation of an alloy of that composition, while those that also contain B or C typically take a very anisotropic, layered structure. As an example, \ce{Re3Ir} highlighted in \cref{fig:single-prop} can be interpreted as an ordered approximation of a \ce{Ir_{x}Re_{1-x}} alloy, which has been previously synthesized and is known to exist in solid solution \cite{gromilov2013synthesis}. The \ce{Re3B2C} example is more unusual, with layers of all Re, B, Re, C, Re, ..., and nominally can charge balance. We note that the bulk modulus calculated is an averaged quantity over the full elastic tensor, and we have not examined the directional bulk moduli in these highly anisotropic systems.

\subsection{Designing low-supply-chain risk magnets}
\label{sec:supp_multi-prop}

This section provides supplementary information to the results in \cref{sec:multi-prop}.

\subsubsection{Additional experimental details}
To generate structures conditioned on magnetic density and HHI score, we fine-tune our base model on these two properties, encoded as in \cref{sec:supp-single-prop}.
To evaluate the performance of our model, we proceed as detailed in \cref{sec:supp-single-prop}, and generate 512 samples with our fine-tuned model, by conditioning on a magnetic density value of 0.2 $\text{\AA}^{-3}$ and an \gls{HHI} score of 1200.
Of those, 130 samples remain after filtering by stability and uniqueness following the \gls{DFT} relaxation. 
Finally, a total of 112 structures pass the novelty check w.r.t. the reference dataset and are reported in \cref{fig:multi-prop}(a).

\subsubsection{Additional qualitative analysis of structures}\label{subsec:supp_multi-prop_analysis}

Targeting a low \gls{HHI} index in addition to high magnetic density steers \mbox{MatterGen} away from generating structures with Co, which is associated with poor \gls{HHI} scores. Example structures include \ce{Fe_x Mn_{1-x} O} rocksalt alloys (\ce{MnFe3O4}, \ce{MnFe8O9}); however, they only exhibit a high magnetization density in a hypothetical ferromagnetic state, and not in their actual antiferromagnetic ground state. Other similar example outputs include a defected FeO containing vacancies (\ce{Fe8O9}), and a body-centered-cubic \ce{Fe8Au} system, both of which are well-known experimentally. While the joint optimization task could further be extended to produce more reasonable candidates---e.g., by penalizing expensive elements, and prefering metallic systems more likely to be ferromagnetic---the overall performance of the model with respect to the labels used for training is reasonable. It is possible that a better treatment of alloy systems would be required to improve performance of the high magnetic density generation task, and to ensure that the generated structures are truly novel.

\end{appendices}

\bibliography{sn-bibliography}

\end{document}